\documentclass[aps,prl,twocolumn,letterpaper,superscriptaddress,showpacs]{revtex4-1}
\usepackage{graphicx}
\usepackage{CJK}
\usepackage{verbatim}
\usepackage{lineno}
\usepackage{bm}
\usepackage{mathptmx}
\usepackage{subfigure}
\usepackage{amsmath}
\makeatletter

\usepackage{titlesec}
\titleformat*{\section}{\large\centering\bfseries}
\usepackage{subfiles}
\usepackage{natbib}
\usepackage{graphicx}
\usepackage{amsmath}
\usepackage{lipsum}

\begin{document}

\begin{CJK*}{UTF8}{bsmi}

\title{Probing a Bose Metal via Electrons: \\
Inescapable non-Fermi liquid scattering and pseudogap physics}

\author{Xinlei Yue (\CJKfamily{gbsn}岳辛磊)}
\affiliation{Zhiyuan College, Shanghai Jiao Tong University, Shanghai 200240, China}
\affiliation{Tsung-Dao Lee Institute, Shanghai 200240, China}
\author{Anthony Hegg}
\affiliation{Tsung-Dao Lee Institute, Shanghai 200240, China}
\author{Xiang Li (\CJKfamily{gbsn}李翔)}
\affiliation{Zhiyuan College, Shanghai Jiao Tong University, Shanghai 200240, China}
\affiliation{Tsung-Dao Lee Institute, Shanghai 200240, China}
\author{Wei Ku (\CJKfamily{bsmi}顧威)}
\altaffiliation{corresponding email: weiku@sjtu.edu.cn}
\affiliation{Tsung-Dao Lee Institute, Shanghai 200240, China}
\affiliation{Key Laboratory of Artificial Structures and Quantum Control (Ministry of Education), Shanghai 200240, China} 
\affiliation{Shanghai Branch, Hefei National Laboratory, Shanghai 201315, China}

\date{\today}

\begin{abstract}
Non-Fermi liquid behavior and pseudogap formation are among the most well-known examples of exotic spectral features observed in several strongly correlated materials such as the hole-doped cuprates, nickelates, iridates, ruthenates, ferropnictides, doped Mott organics, transition metal dichalcogenides, heavy fermions, $d$- and $f$-electron metals, etc. We demonstrate that these features are inevitable consequences when fermions couple to an unconventional Bose metal~\cite{hegg2021bose} mean field consisting of lower-dimensional coherence. Not only do we find both exotic phenomena, but also a host of other features that have been observed e.g. in the cuprates including nodal anti-nodal dichotomy and pseudogap asymmetry(symmetry) in momentum(real) space. Obtaining these exotic and heretofore mysterious phenomena via a mean field offers a simple, universal, and therefore widely applicable explanation for their ubiquitous empirical appearance.
\end{abstract}
\maketitle
\end{CJK*}

\maketitle

Strongly correlated materials present some of the most mystifying phenomena in physics. The reason for this is two-fold. First, these materials have a tendency toward exotic properties that defy conventional theory. Second, and perhaps more importantly, many of these exotic observations exhibit universality in that they can be found in a range of materials that seem completely unrelated at the microscopic level. The combination of their exotic nature and the mathematical difficulty in theoretical treatment has generated an enormous amount of research~\cite{dagotto1994correlated,damascelli2003angle,lee2014amperean,battisti2017universality}
%{anisimov1991band,dagotto1994correlated,imada1998metal,nagaosa1999quantum,tsuei2000pairing,damascelli2003angle,lee2006doping,stewart2011superconductivity}
over many decades with little sign of stagnation, and this research has been a wellspring of novel and exciting physics. Yet, comparatively little progress has been made in producing theory that measures up to the empirical universality.

Within electronic spectral observations, non-Fermi liquid (NFL) scattering and pseudogap (PG) formation are two such exotic and universal phenomena. A phenomenon is labeled as NFL scattering when the corresponding scattering rate of quasiparticle excitation violates $T^2$ and $\omega^2$ dependence~\cite{abrikosov} near the chemical potential. NFL scattering has been observed in cuprates~\cite{valla1999evidence,abdel2006anisotropic,johnston2012evidence,kaminski2005momentum}, ruthenates~\cite{bruin2013similarity}, ferropnictides~\cite{fink2015non,dai2015spin}, transition metal dichalcogenides~\cite{umemoto2019pseudogap}, and heavy fermions~\cite{lohneysen1994non,paglione2007incoherent}. A PG is said to form when a gap scale appears in the quasiparticle excitation spectrum, but a significant number of states fill the gap even at low temperature. PG behavior is widely found in cuprates~\cite{tao1997observation,renner1998pseudogap,timusk1999pseudogap,lang2002imaging,damascelli2003angle,yang2008emergence,particlehole,pseudogap_t_depen,he2011single,arpes_cuprate,arpes_cuprate_2,he2011single},
%~\cite{tao1997observation,renner1998pseudogap,timusk1999pseudogap,lang2002imaging,damascelli2003angle,yang2008emergence,particlehole,pseudogap_t_depen,he2011single,arpes_cuprate,arpes_cuprate_2,arpes_cuprate_3,he2011single}
nickelates~\cite{nickelate}, irridates~\cite{kim}, ruthenates~\cite{iwaya2007local}, ferropnictides~\cite{ironpnictides_1,ironbase_1}, Mott organics~\cite{powell2011quantum}, and transition metal dichalcogenides~\cite{chen2017emergence,umemoto2019pseudogap,borisenko2008pseudogap}.

The names of these two phenomena highlight both the violation of and the empirical universality they share with the corresponding conventional theory. In a Fermi liquid (FL) the vanishing phase space near the chemical potential suppresses scattering and preserves the quasiparticle nature. The resulting energy and temperature dependence of the single particle spectrum is highly constrained and very robust. As for gap formation, perfectly coherent scattering (from a mean field for example) always opens a clean gap. As long as the gap scale remains dominant, states only fill it at best exponentially slowly. The robust universal nature of these conventional phenomena strongly suggests that their violation would be difficult and rare, yet empirical evidence directly contradicts this picture. Instead, NFL and PG phenomena have been found in many seemingly unrelated materials covering a large region of their phase diagrams including at the zero temperature limit, suggesting that they harbor an as-yet unidentified robust universal quantum nature of their own.
%, in analogy with the conventional physics above.

Existing theoretical treatments of NFL behavior typically start with a FL and enhance scattering channels at low energy to disrupt the quasiparticle nature. Several examples include marginal Fermi liquid~\cite{varma1989phenomenology}, multichannel Kondo physics~\cite{cox1996two},
%~\cite{nozieres1980kondo,schlottmann1993multichannel,cox1996two,cox1998exotic}
orbital fluctuations~\cite{lee2012non}, holographic methods~\cite{iqbal2012lectures}, and quantum critical fluctuations~\cite{millis1993effect,castellani1996non,abanov2000spin,oganesyan2001quantum,lohneysen2007fermi,metlitski2010quantum,abrahams2012critical,isobe2016emergent,moriya2012spin,liu2018itinerant,lee2018recent}.
%~\cite{hertz1976quantum,moriya1973effect,moriya1973effect2,millis1993effect,castellani1996non,abanov2000spin,oganesyan2001quantum,abanov2004anomalous,rech2006quantum,sachdev2007quantum,lohneysen2007fermi,lee2009low,metlitski2010quantum,metlitski2010quantum1,abrahams2012critical,sur2015quasilocal,isobe2016emergent,lederer2017superconductivity,moriya2012spin,liu2018itinerant,lee2018recent}
However, since a systematic theoretical treatment of unconventional features over a wide range of experiments is lacking, a correspondingly systematic understanding of these relatively universal characteristics remains elusive.

Theories exhibiting PG behavior typically start from the formation of a clean gap and then introduce long-range fluctuations to refill it. Some examples include preformed pairs scenarios~\cite{loktev2000phase,loktev2001phase} and models with spin or charge correlations~\cite{schmalian1999microscopic,sdw,sedrakyan2010pseudogap,PhysRevLett.31.462,kuchinskii2012electronic,PhysRevLett.108.186405,PhysRevB.83.014514}. These methods rely on coherent order parameters and then introduce fluctuations to weaken them so they vanish at long distances. Such fluctuations require proximity to a quantum critical point or phase boundary, but experiments often find PG formation over wide parameter ranges even at very low temperatures~\cite{kordyuk2015pseudogap}. Perhaps these scenarios could be realized in nature, but such restrictive constraints eliminate any chance to account for the observed universality yet again.

Since the theories above are highly specialized to exhibit one exotic phenomenon each, it is perhaps not surprising that NFL and PG theoretical understanding have essentially no overlap. Describing one exotic phenomenon in one material is difficult enough, let alone two seemingly unrelated such mysteries. Yet many materials exhibit both NFL scattering and PG formation in the same experimental parameter range. They are also both properties of the single particle excitation spectrum, so when they occur simultaneously over a broad region of the phase diagram of a material, it is natural to assume a universal mechanism would exhibit both phenomena.

Here we identify a robust paradigm where the exotic phenomena of non-Fermi liquid scattering rates and pseudogap formation inevitably result. The crucial component is an unconventional effective mean field derived from a recently proposed Bose metal~\cite{hegg2021bose}. The mean field studied here only exhibits lower-dimensional coherence leaving it a measure of incoherence, and it is this incoherence that defeats the conventional framework. To demonstrate the qualitative effects of this idea, we couple this mean field to simple non-interacting fermions so that all the characteristic features originate only from the mean field. Not only do we find NFL and PG behavior but also a range of additional phenomena all of which have been observed together in nature~\cite{arpes_cuprate,arpes_cuprate_2,particlehole,tao1997observation,renner1998pseudogap,lang2002imaging}. Producing these phenomena from a single mean field provides the framework for a systematic comparison of these unconventional features and contains the wide applicability necessary to account for existing and future observations.

%Figure 1
\begin{figure}[]
\centering
\subfigure{
\begin{minipage}[c]{0.45\linewidth}
\centering
\includegraphics[width=0.98\textwidth]{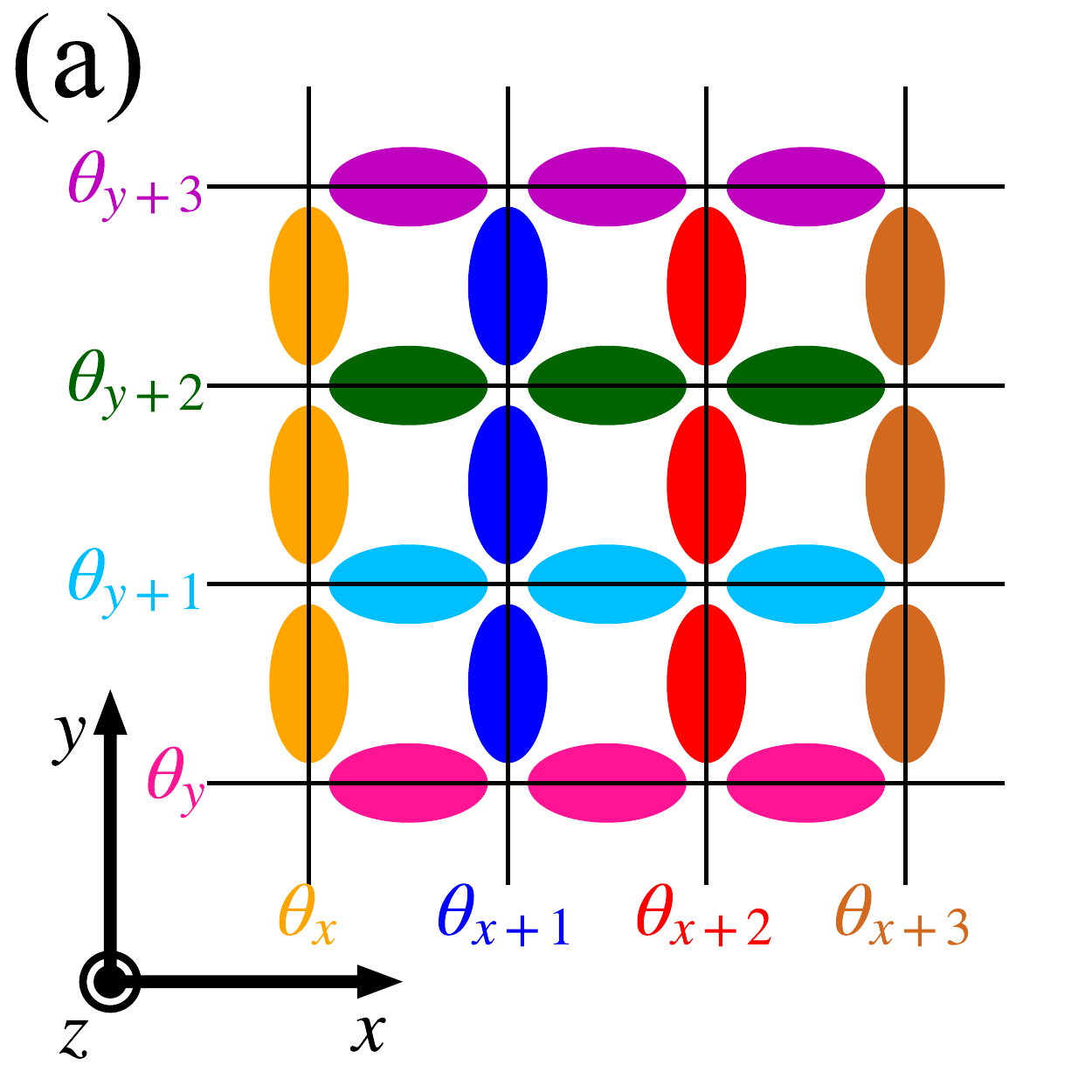}
\end{minipage}
%\label{meanfield}
}
\subfigure{
\begin{minipage}[c]{0.5\linewidth}
\centering
\includegraphics[width=0.98\textwidth]{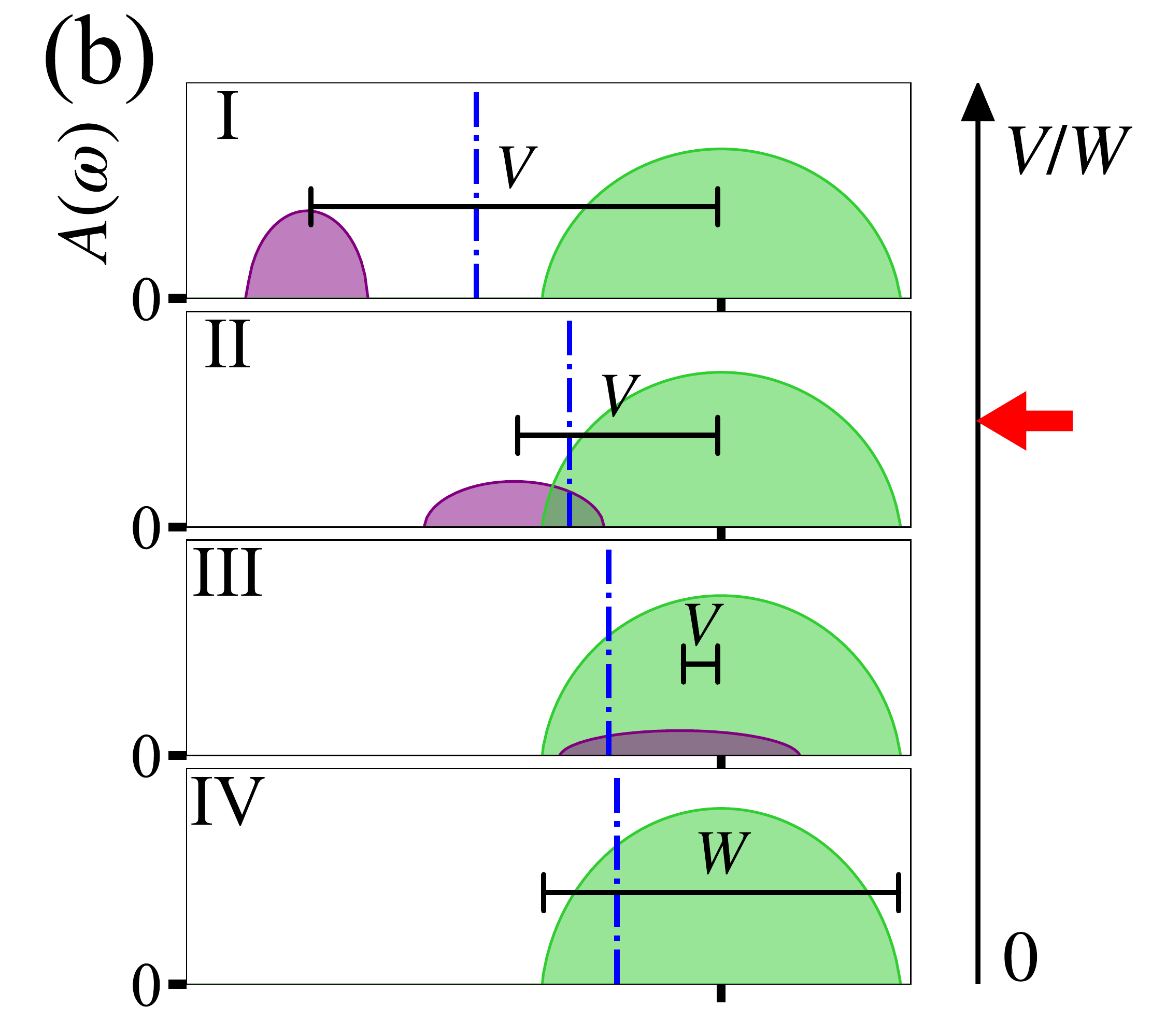}
\end{minipage}
%\label{quasiboundschema}
}
\caption{(a) Illustration of the Bose metal one-body phase structure (XY plane). The bond-centered bosons, shown by ellipses, form a checkerboard lattice. Ellipse color indicates dominant local phase of the bosons. Each color represents a fixed randomly chosen phase factor. (b) Illustration of the spectral weight for several ratios of binding energy with bandwidth. Unbound and bound quasiparticle density of states (DOS) are given in green and purple respectively. The chemical potential is marked by blue dotted lines. The region we study is represented by panel II marked with a red arrow. Comparison could be made with the numerical results~\cite{kuleeva2014normal}.}
\label{illu}
\end{figure}

%Figure 2
\begin{figure*}
	\includegraphics[width=0.98\textwidth]{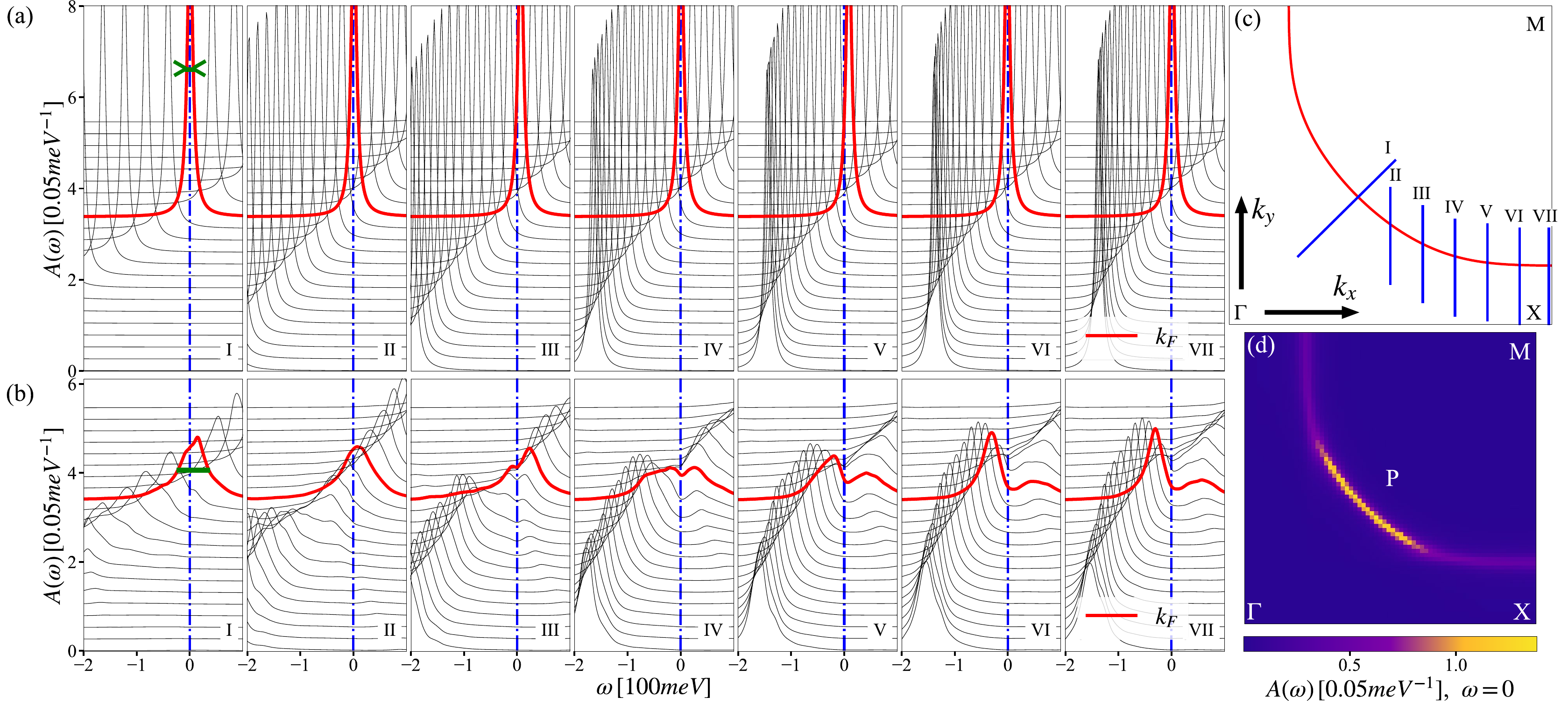}
	\caption{Spectral function $A(\mathbf{k},\omega)$ at doping $\delta\approx0.11$ (a) without and (b) with coupling to the Bose metal for selected lines in $\mathbf{k}$-space denoted in (c). Red lines indicate the momentum nearest to the Fermi vector $\mathbf{k}_\mathrm{F}$. (d) the spectral function near the chemical potential reveals 'Fermi arc' structure.}
	\label{Fig_1}
\end{figure*}

To generate NFL and PG behavior we must break the highly constrained nature of Fermi quasiparticles. The former by opening up a significant scattering channel at low-energy and low-temperature. The latter by coupling to incomplete coherent structure. These conditions can be met by coupling to a recently proposed Bose metal~\cite{hegg2021bose}. This Bose metal is a stable phase of matter that is neither superfluid nor insulator and it is comprised of so-called type-I (real, conserved) bosons\cite{KohnBosons}, so it is not confined to the vicinity of any critical transition as required by type-II bosons (excitations such as phonons, magnons, etc.). In one example, the ground state of the Bose metal consists of coherent 2D planes that are effectively independent of each other to one-body level in the low-energy limit. This should be contrasted with the typical long-range isotropic coherence found in, for example, superfluidity. The dominant one-body phase structure for this example is illustrated in Fig.\ref{illu} (a) in which each coherent region is represented by a single color.

We demonstrate the effects of scattering against our Bose metal by coupling it to a simple tight binding Fermi gas

\begin{equation}
    H=H_{\text{kinetic}}+H_{\text{scattering}}
\end{equation}
 where $H_\text{kinetic}$ contains the bare fermion dispersion and $H_\text{scattering}$ describes the coupling between these fermions to the dominant bosons in the Bose metal, chosen to be Bogoliubov-like

\begin{equation}
\label{coupEq}
H_{\text{scattering}}=\sum_{(i,j) \in N N}  \tilde{V} b_{ij}^{\dagger}c_{j} c_{i} + h.c.
\end{equation}
Here $\tilde{V}$ is the effective coupling strength between fermion $c^\dagger_i$ at site $i$ and $j$ and boson $b^\dagger_{ij}$ located at the bond between them. $NN$ denotes nearest neighbors sites $i$ and $j$ in the $XY$-plane. This choice for the coupling is not required, but it represents a scenario~\cite{yildirim2011kinetics} where the bosons are formed from binding the fermions into local pairs at high energy and are therefore indistinguishable from each other. Suppose we are in the regime as shown by the red arrow in Fig.\ref{illu} (b) where most of the fermion weight is in a tightly bound state (in purple) represented by a boson, yet some weight remains in the unbound fermionic state (in green). Assuming the boson density is much higher than that of the fermions, the influence of the fermions on the Bose metal structure is negligible. For convenience, we further assume a scale $\tilde{V} \sim t$ of the order of the nearest neighbor hopping strength $t$.

Since the dominant bosonic energy scale is very low compared to the fermionic bandwidth, we replace our Bose metal with a well-justified (c.f. \textbf{Appendix A}) static effective mean field at low temperature

\begin{equation}
\label{HScatt}
\begin{aligned}
H_{\text{scattering}} \rightarrow \sum_{k_x k_y} \Big\{&\sum_{q}V_y(q)c_{k_x,k_y}c_{-k_x,-k_y+q}\\
+&\sum_{q}V_x(q)c_{k_x,k_y}c_{-k_x+q,-k_y}\Big\}+h.c.,
\end{aligned}
\end{equation}
where the notation $k_{z}=0$ has been dropped given the perfect in-phase coherence along the $z$-direction in our Bose metal structure shown in Fig.\ref{illu}(a) (c.f. \textbf{Appendix B}). As shown below, the resulting spectral features are very broad in energy and momentum and therefore insensitive to fine-tuned parameters.

%Figure 3

\begin{figure}[h]
	\renewcommand\thefigure{\arabic{figure}}
	\includegraphics[width = 0.9\linewidth]{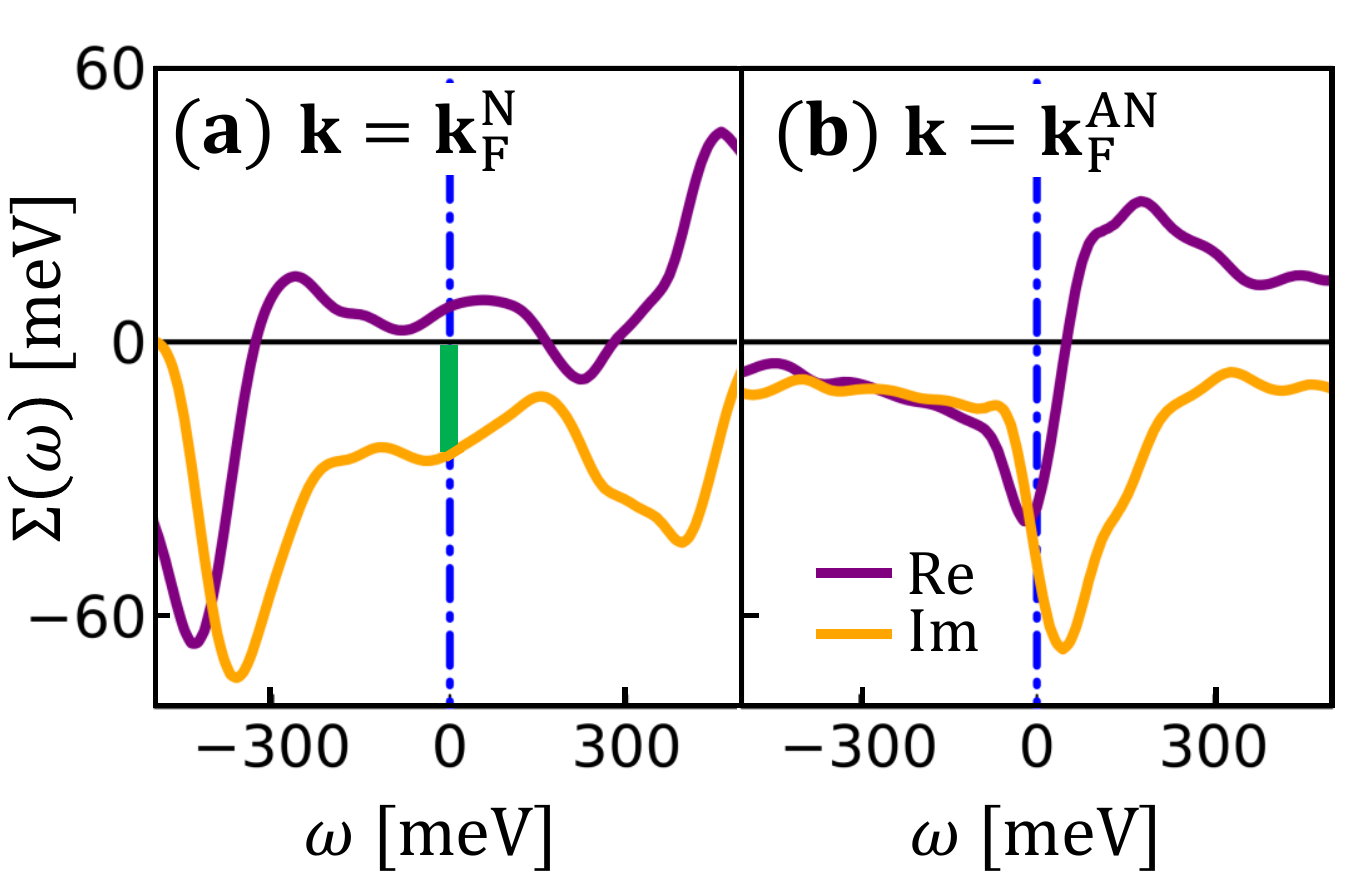}
	\caption{The self energy $\Sigma(\omega)$ at nodal (a) and antinodal (b) Fermi momenta $\mathbf{k}_\mathrm{F}^\mathrm{N}$ and $\mathbf{k}_\mathrm{F}^\mathrm{AN}$ respectively. Notice the unusual finite imaginary part (highlighted by the green line) at $\omega=0$ for $\mathbf{k}=\mathbf{k}_\mathrm{F}^\mathrm{N}$ that produces the non-Fermi liquid scattering rate even at zero temperature. In contrast, for $\mathbf{k}_\mathrm{F}^\mathrm{AN}$ a peak near $\omega=0$ pushes open the pseudogap.}
	\label{self-energy}
\end{figure}

%Summarize kep points of method and results (Fig. 2 info)

Fig.\ref{Fig_1} summarizes the main features of our results using the tight binding parameters of the cuprates~\cite{he2011single} as an example. Panels (a) and (b) correspond to the momentum-diagonal element of our resulting spectral functions $A(\mathbf{k}, \omega) = -\frac{1}{\pi}\mathrm{Im}\;G(\mathbf{k}, \mathbf{k^\prime}=\mathbf{k},\omega)$ along the momentum cuts I-VII illustrated in panel (c) before and after coupling to the Bose metal respectively. The fermionic chemical potential is chosen separately for each case in order to fix the particle number. Panel (d) gives the momentum dependent spectral weight at the chemical potential, which clearly shows a dichotomy between the 'nodal' region near P$=(\pi/2,\pi/2)$ and the 'anti-nodal' region near X$=(\pi,0)$ reminiscent of the infamous 'Fermi arc' phenomenon\footnote{Figure \ref{Fig_1}(d) already resembles the experimental Fermi arc, although there is some debate~\cite{yang2011reconstructed} on the shape at the tip of the arc. In our theory, this inessential detail is strongly dependent on the choice of bare fermionic structure, and such features could easily be accounted for by using a more 'first-principles' system incorporating, for example, strong magnetic correlations of higher energy scale.}.

Near the nodal region (panels I-III) $A(\omega)$ acquires a rather large scattering rate (width denoted by green bar in I) near the chemical potential even at the Fermi wave vector $k_{F}$ (red curve). In a clear departure from Fermi liquid theory, when mapped to the self-energy $\text{Im}\Sigma(\mathbf{k},\omega;T)$ this corresponds to a finite imaginary part at the chemical potential (c.f. Fig.~\ref{self-energy}). Having a finite scattering rate at the chemical potential even in the low-temperature limit is of profound significance. This implies anomalous dissipation in a clean quantum system of fermions, unimaginable within the standard lore. 

\begin{figure}[ht]
	\includegraphics[width = 0.65\linewidth]{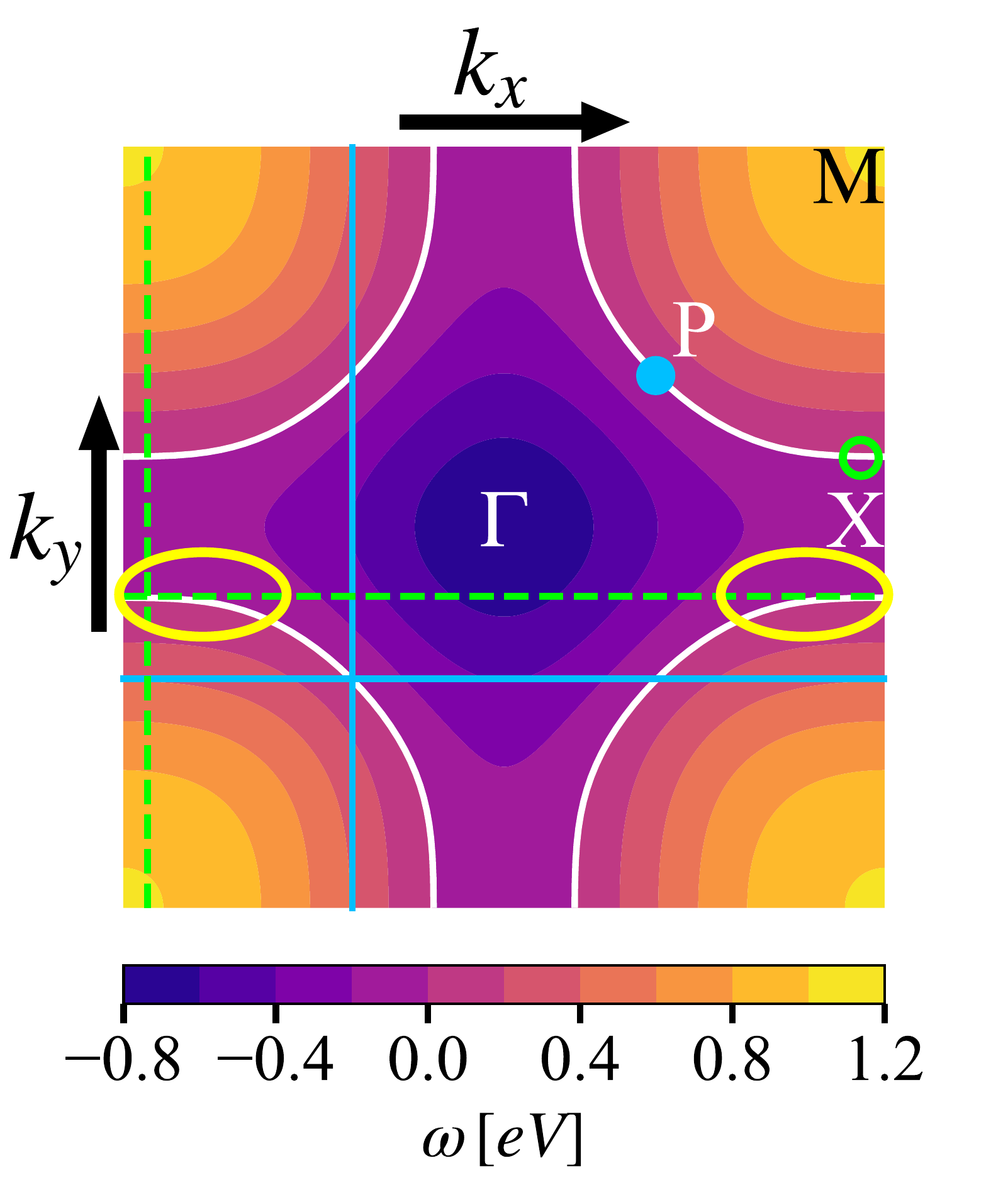}
	\caption{Energy contours of the band structure at $\delta \approx 0.11$ doping. The white line indicates the Fermi surface. States with crystal momenta near P$=(\pi/2,\pi/2$) (X=$(\pi,0)$), marked by the green circle (blue disk), couple with points along the green dashed lines(blue solid lines) through the Bose metal mean field. This large coupling space leads to scattering in general broadening all peaks, but the coupling near X involves many states in a small energy window (near a van Hove singularity) resulting in scattering so strong the quasiparticle peaks are destroyed completely.}
	\label{contour}
\end{figure}

The origin of this large scattering rate can be understood clearly from the structure of the Bose metal effective mean field coupling as illustrated in Fig.\ref{contour}. Due to the lack of translational symmetry of the Bose metal mean field, each boson couples a fermion state into every state along a line. For example, a state at P would couple to every state along the vertical and horizontal blue lines through scattering against bosons that move horizontally and vertically respectively in Fig.\ref{illu}(a). This significantly opens up the phase space for scattering as compared to bosons with well-defined momenta. Since the states along the blue lines cover a broad range of energy, the fermionic poles are always broadened even down to the chemical potential. This enhanced phase space for scattering is even more effective than the typical NFL scenario, which involves scattering against bosons of well-defined momentum~\cite{jiang2019non,zhang2021infinity,lake2021bose}. Furthermore, since the Bose metal is a stable phase composed of real bosons (with fixed density), the effect of coupling is actually strengthened in the low-temperature  limit or far away from a phase boundary.

It is important to note that the fixed boson number in our theory is essential to sustain large scattering in the low-temperature limit. In contrast, the typical scenario in which fermions scatter against thermally excited bosons (e.g. phonons, magnons, spinons, holons, phasons, acoustic plasmons, orbitons, excitons, etc.) finds diminishing scattering at low-temperature. This limitation forces the existing attempts to account NFL behavior to resort to critical fluctuations, which allow for a macroscopic number of critical bosons~\cite{millis1993effect,lohneysen2007fermi,moriya2012spin,lee2018recent}.
%~\cite{hertz1976quantum,millis1993effect,moriya1973effect,moriya1973effect2,lohneysen2007fermi,moriya2012spin,lee2018recent}
However, this overconstrained scenario is inconsistent with observations of NFL behavior over a broad region of the phase diagram, for example in the pseudogap regime of the cuprates, whereas our anomalous dissipation scenario involves a stable phase compatible with these experiments. Incidentally, the stability of the Bose metal as a phase of matter and the qualitative agreement between our results and experiments strongly suggests that the pseudogap regime of the cuprates is a finite temperature phase of a Bose metal.

\begin{figure}[t]
\centering
\includegraphics[width=\columnwidth]{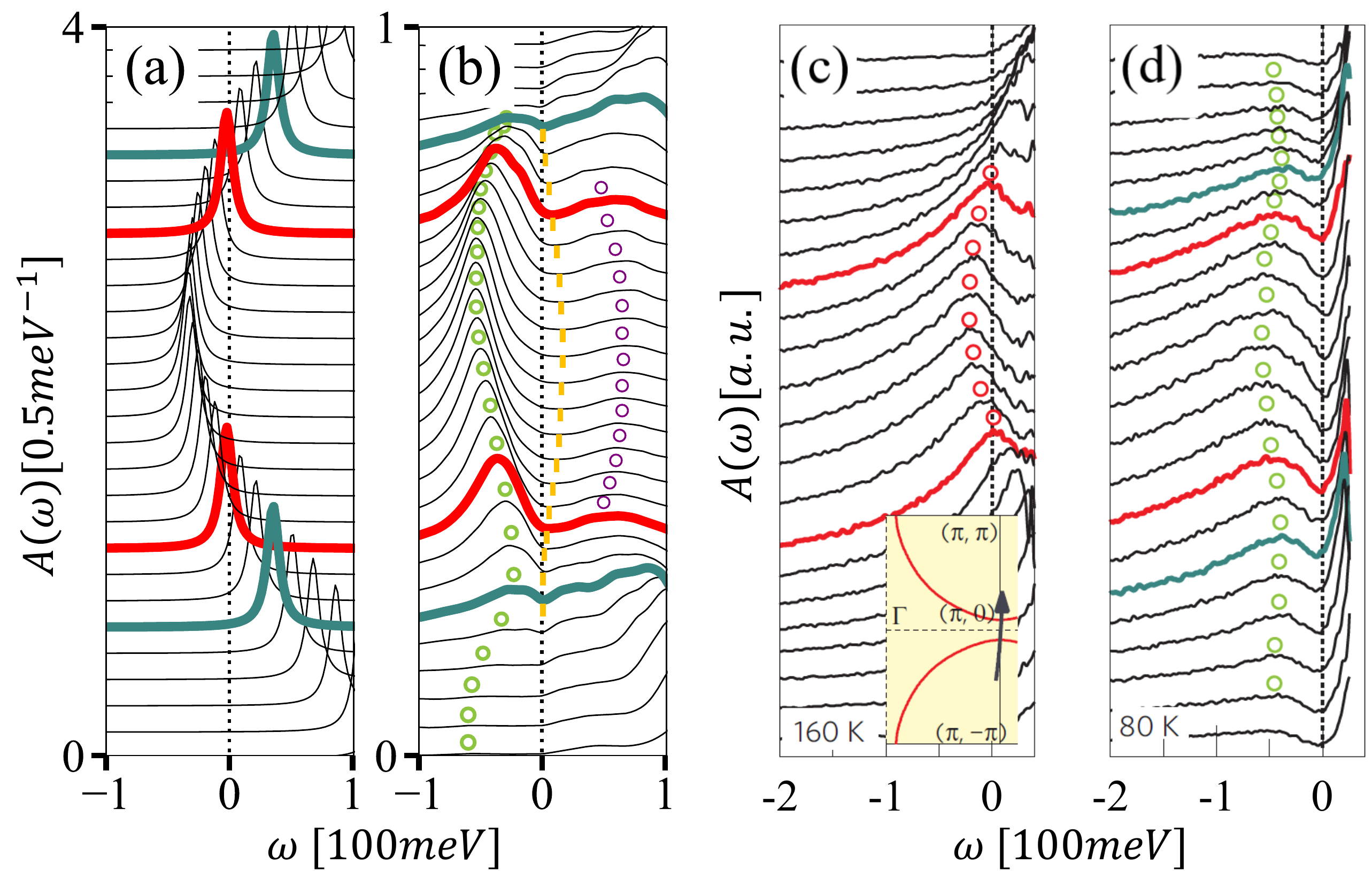}
\caption{Theoretical(a)(b) and experimental(c)(d) spectral functions near the anti-nodal region following the arrow shown in the inset for the high temperature phase(a)(c) and the low-temperature pseudogap regime(b)(d). Experimental plots are inferred from ARPES data for optimally doped Bi2201~\cite{particlehole,arpes_cuprate} ($T_c\approx34K$). Notice that the gap minimum does not occur at the bare Fermi vector(red line), and instead is found(green line) in the back-bending band(green circles) generated through the formation of the pseudogap. Also notice the dispersion of the gap center (yellow vertical bars) defined by the minimum of $A(\omega)$ in the gap.}
\label{detailed_arpes}
\end{figure}

Our large scattering rate at zero temperature even right at the chemical potential is unimaginable to-date since many theories~\cite{varma1989phenomenology,millis1993effect,castellani1996non,abanov2000spin,oganesyan2001quantum,lohneysen2007fermi,metlitski2010quantum,abrahams2012critical,isobe2016emergent,moriya2012spin,liu2018itinerant,lee2018recent,leong2017power} have willingly explored extreme exotic fermions but strictly maintain dissipationless behavior when the system is clean (disorder-free). On the other hand, having accepted this new possibility, any smooth analytic scattering rate can show NFL behavior $\text{Im}\Sigma(\omega,T) \approx c + a \omega + b T + \cdots$ (c.f. Fig.~\ref{self-energy}). Incidentally, experiments~\cite{valla1999evidence,fink2015non} observe the same smooth analytic behavior that is always finite at low-energy, which only transforms into an exotic non-analytic mystery if $c$ is assumed to arise from disorder~\cite{abrahams2000angle} and subsequently removed.

Moving away from the nodal region toward the anti-nodal region (Fig.\ref{Fig_1} V-VII) we find qualitatively different behavior. Here at low-energy $A(\omega)$ loses its quasiparticle peak-like structure. Instead, we find wide but distinct peaks above and below the chemical potential defining a clear gap scale. Notice that there is substantial weight within this gap, a feature typically regarded as a pseudogap.

Despite the dichotomy between nodal region near P and the anti-nodal region near X, our pseudogap can be understood by examining the coupling to the Bose metal mean field once again. In both regions we couple to states over a broad energy range and the difference between these regions changes very smoothly. But the coupling along the green dotted lines in Fig.\ref{contour} now includes quantitatively more states of similar energy circled in yellow, which makes the coupling effective enough to open a gap scale. However, the incomplete coherence of our mean field cause the gap to be filled with states as opposed to the case where complete coherence opens a clean gap.

\begin{figure}[t]
\centering
\includegraphics[width=\columnwidth]{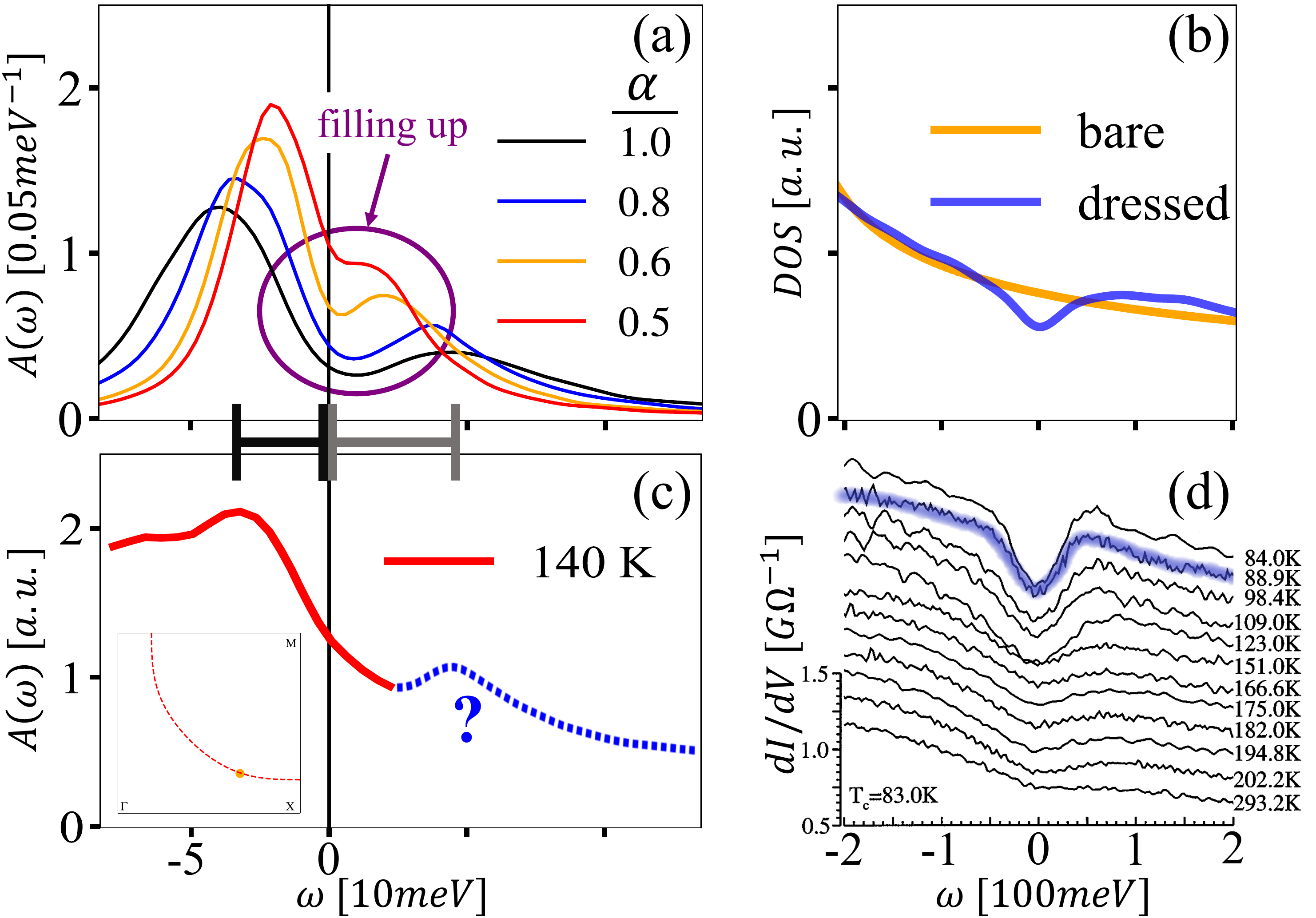}
\caption{Seemingly contradictory experimental spectral functions of underdoped BSCCO with asymmetric/symmetric pseudogap structure in (c) momentum (spectral function at the end of the fermi arc, shown by the orange dot in the inset. Data reproduced from~\cite{yang2008emergence}) and (d) real~\cite{renner1998pseudogap} space. Similar features are observed to be compatible in our theoretical results (a) and (b). The effect of thermal reduction $\alpha=\Phi(T)/\Phi(0)$ of the order parameter($\Phi$) is shown in panel (a) as well displaying characteristics of gap filling as opposed to gap closing.}
\label{exptcomp}
\end{figure}

Surprisingly, our pseudogap exhibits many of the symptoms typically associated with pseudogap observations. In Fig.\ref{detailed_arpes} we exhibit a back-bending band traced in green circles and accompanied by a shadow band traced in purple circles in panel (b) in agreement with experiments in panel (d). The back-bending band and the gap center between these bands (outlined in vertical yellow bars) develop dispersion~\cite{particlehole,arpes_cuprate,yang2008emergence,lee2014amperean}, and correspondingly $k_F$ (red line) is no longer the gap minimum (green lines). The diagnosis is clear: we couple to many states of different energy thereby breaking one of the conditions required to make $k_F$ the gap minimum.

A more important characteristic of our pseudogap is how it dissolves with increasing temperature. Fig.\ref{exptcomp}(a) illustrates that upon thermal reduction of the mean field by a factor $\alpha=\Phi(T)/\Phi(0) \sim 0.5$, the gap gets filled up by more states faster than it closes in agreement with experiments~\cite{pseudogap_t_depen,particlehole,arpes_cuprate}. Our pseudogap becomes indistinguishable from the continuum at a temperature $T^{*}$ long before the Bose metal mean field (gap scale) is depleted at a higher temperature $T_{\text{BM}}$.

Perhaps the most significant feature of our pseudogap, as illustrated in Fig.\ref{exptcomp}(a), is a strong asymmetry of the peak energies above and below the Fermi energy. This asymmetry is also observed experimentally as in Fig.\ref{exptcomp}(c) and has led to preference toward fluctuating charge density wave, spin density wave~\cite{schmalian1999microscopic,sdw,sedrakyan2010pseudogap} or pairing density wave~\cite{lee2014amperean,dai2020modeling} over a Bogoliubov (preformed pair). However, we were able to obtain this asymmetry using a Bogoliubov coupling since our Bose metal mean field couples fermion states in a line and therefore we do not expect symmetry in general.

Surprisingly, in real space our pseudogap appears to be perfectly symmetric in energy as illustrated in Fig.\ref{exptcomp}(b) despite the strong asymmetry in momentum space considering the former is the sum over the latter. It is precisely this sum that loses the information about partial coherence and leaves only the symmetry of our Bogoliubov coupling. In fact, this apparent contradiction between real and momentum space has been haunting the experimental community in the cuprates for decades between scanning tunneling microscopy (STM)~\cite{renner1998pseudogap,lang2002imaging} and angle-resolved photoemission spectroscopy (ARPES) experiments. To our knowledge, our theory provides the first natural resolution to this big puzzle.

In essence, our theory is successful because it encapsulates simple ingredients required for NFL scattering rate and PG phenomena. Through the interplay between coherence and incoherence of the one-body phase in the quantum many-body states, we introduce just enough scattering to generate a \textit{finite} non-Fermi liquid scattering rate even right at the chemical potential at zero temperature. On top of that, when assisted by a large amount of phase space, for example near a van Hove singularity, the system will further generate a pseudogap due to the combination of partial coherence (gap scale) and incoherence (states inside). To our knowledge, no theory prior to ours has managed to simultaneously generate such a natural NFL scattering rate and the PG in the absence of proximity-induced fluctuations.

It is important to emphasise the quantum nature of our mechanism, in comparison with previous attempts via thermal fluctuation~\cite{schmalian1999microscopic,sdw,sedrakyan2010pseudogap,PhysRevLett.31.462,kuchinskii2012electronic}.
Our work exploits the unique imperfect phase coherence of the \textit{quantum} Bose metal state. The bosons that scatter the electrons are not excitations of the system, but instead emerged particles (for example tightly bound state of two fermionic carriers) that constitute the Bose metal ground state\cite{hegg2021bose}. The resulting exotic phenomena therefore persist to zero-temperature limit. In great contrast, the previous attempts ~\cite{schmalian1999microscopic,sdw,sedrakyan2010pseudogap,PhysRevLett.31.462,kuchinskii2012electronic} made use of soft bosonic excitation modes that are only effective at temperature higher than their characteristic energy scale. These thermal fluctuations are therefore inactive in the zero-temperature limit (except right above a quantum critical point), unable to account for the experimentally observed large region of the phase diagram that extends to zero temperature.

For those interested in the cuprates, our successful reproduction of NFL scattering, PG formation, and many related symptoms found in the cuprates leads one to wonder whether the pseudogap \textit{regime} of the cuprate phase diagram below $T^*$ is actually contained within the finite temperature \textit{phase} of a Bose metal below a higher temperature $T_\text{BM}$. In fact, the Bose metal mean field used here directly results from the phase frustrated regime~\cite{yildirim2015weak,hegg2021bose} of the emergent Bose liquid theory~\cite{yildirim2011kinetics, lang2019mottness} of the cuprates, which has been demonstrated to describe many physical properties simultaneously with the \textit{same} model and a single \textit{fixed} set of parameters (c.f. Appendix C). Within this EBL model, it is not surprising that the observations from single particle fermion probes such as ARPES and STM are so difficult to explain. Such experiments would be probing a probe (dilute fermion system) of the dominant EBL physical state (a Bose metal) and therefore adding an extra barrier to trace from observation to microscopic picture.

In conclusion, we obtain non-Fermi liquid scattering and pseudogap formation by coupling a simple tight binding Fermi gas to an effective mean field for a recently proposed Bose metal with one-body sub-dimensional coherence. This serendipitous result generates a new paradigm where a generic fermionic system couples to a generic real bosonic field of incomplete coherence to produce unconventional physics. Relying on real bosons that do not vanish in the low temperature limit, we find a finite scattering rate at zero frequency and zero temperature. It is this non-trivial scattering rate at the chemical potential that is ultimately responsible for both non-Fermi liquid scattering and pseudogap formation. This coupling produces additional qualitative characteristics including nodal anti-nodal dichotomy (resulting in Fermi arcs) and particle-hole asymmetry(symmetry) in momentum(real) space as well as other interesting features consistent with experiments. A beneficial byproduct of our study is that it appears to suggest an explanation to the infamous pseudogap of the cuprates. Our method is universal and should therefore be widely applicable to many materials exhibiting these exotic phenomena.

\begin{acknowledgments}
We thank Dong Qian for discussion regarding ARPES experiments. This work is supported by National Natural Science
Foundation of China (NSFC) under Grants No. 12274287 and 12042507, and Innovation Program for Quantum Science and
Technology No. 2021ZD0301900
\end{acknowledgments}

%\onecolumngrid
\section{Appendix}
\author{}
\maketitle
\renewcommand{\theequation}{A.\arabic{equation}}
\subsection{A. Justification of mean-field approximation}

Here we justify the use of the effective mean-field approximation in our study. The key points are that the influence of slow bosonic modes on the typical dynamics of fast fermionic modes can be well-approximated by a static field with an averaged momentum $p_0$.
As explained in the text, we treat the bosons as approximately unaffected by the fermions because of the relatively small population of the latter. The effect onto the fermionic system can be expressed via
\begin{equation}
    H_f=H_0+H_{\text{coupling}}.
\end{equation}
Here $H_0$ is the fermion Hamiltonian (for example a simple tight-binding Fermi gas) modified by coupling to the predefined bosonic modes $b^\dagger_p$ of lattice momentum $p$ via
\begin{equation}
\label{SuppBogo}
    H_{\text{coupling}} = \sum_{\tilde{p},p}\Tilde{V}_{\tilde{p}}b^\dagger_{\tilde{p}}c_{p}c_{\tilde{p}-p}+\text{h.c.},
\end{equation}
as shown in Fig.A\ref{fig_feymann_1}.
Without loss of generality, we illustrate with a local coupling such that the coupling strength $|V_p| = V$ is $p$-independent.

\begin{figure}[htp!]
\setcounter{figure}{0}
\renewcommand\thefigure{A\arabic{figure}}
\centering  %图片全局居中
\subfigure[]{
\label{fig_feymann_1}
\includegraphics[width=0.18\textwidth]{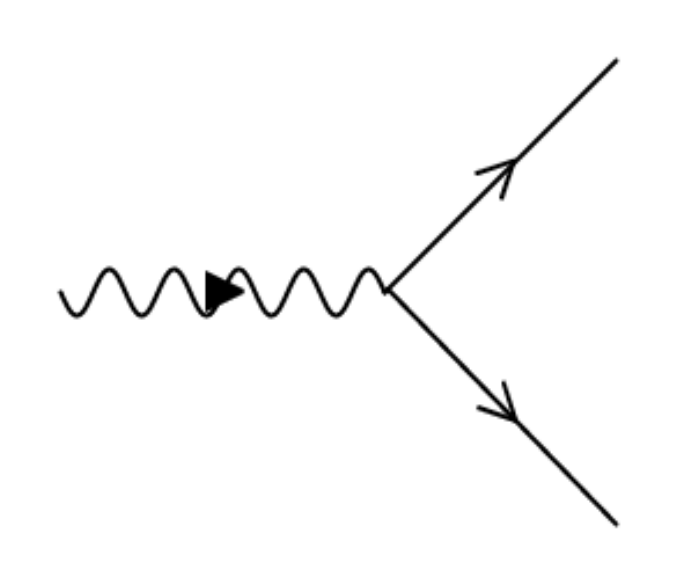}}
\hspace{20mm}
\subfigure[]{
\label{fig_feymann_2}
\includegraphics[width=0.3\textwidth]{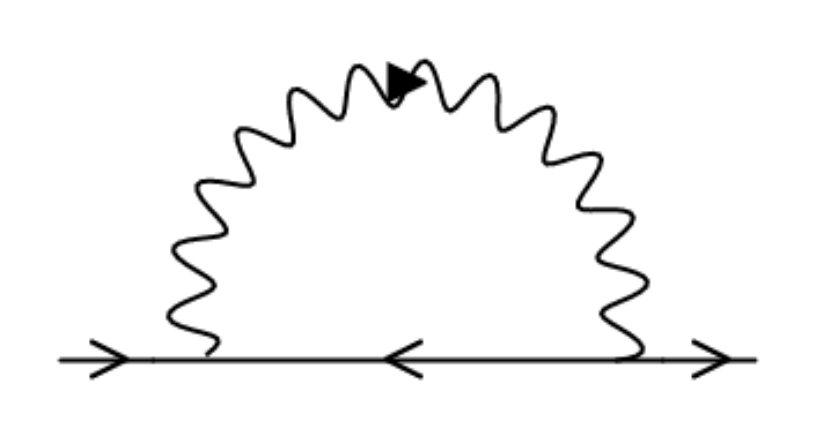}}
\caption{(a) Illustration of the coupling between boson and fermion, here only $b c^\dagger c^\dagger$ is shown. Straight line with single line arrow corresponding to the fermion, the wiggled line corresponding to boson. (b) Illustration of the lowest order correction to the normal green function.}
\label{Fig.main}
\end{figure}

First, it is understandable that objects of very different time scales (frequencies) are insensitive to minute differences due to dispersion.
For example, when the fermionic modes $\xi_p$ are much higher energy than the bosonic modes $\xi_p \gg \epsilon_p \equiv \epsilon_{0} + \Delta$, the scattering of the fermion by such bosons can be approximated as $\epsilon_{p} \approx \epsilon_{0}$.

Further, at small enough temperature, virtually all the bosons are located in a energy window small compared to the bandwidth near the dispersion minimum $p_{0}$. Let $p=p_{0}+\delta$, in this window $\delta<<2\pi$ and the scattering against the bosons will have a momentum transfer dominated by roughly $p_0$, and $\epsilon_{0} \approx 0$. In the presence of condensation this result is even more pronounced.

\begin{figure}[htp!]
\renewcommand\thefigure{A\arabic{figure}}
\centering  %图片全局居中
\subfigure[]{
\label{fig_feymann_3}
\includegraphics[width=0.3\textwidth]{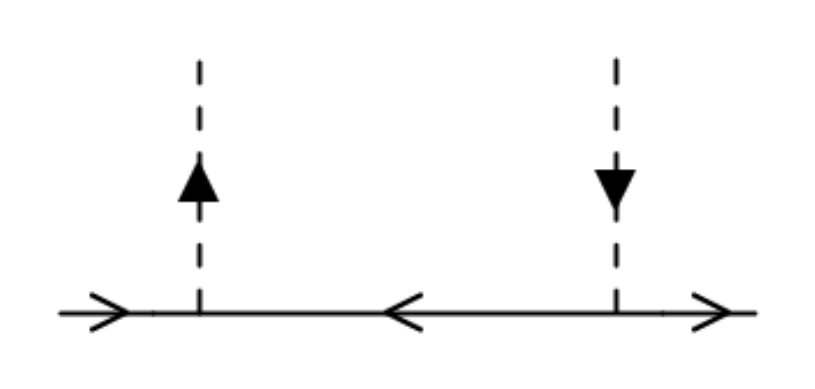}}
\hspace{20mm}
\subfigure[]{
\label{fig_feymann_4}
\includegraphics[width=0.18\textwidth]{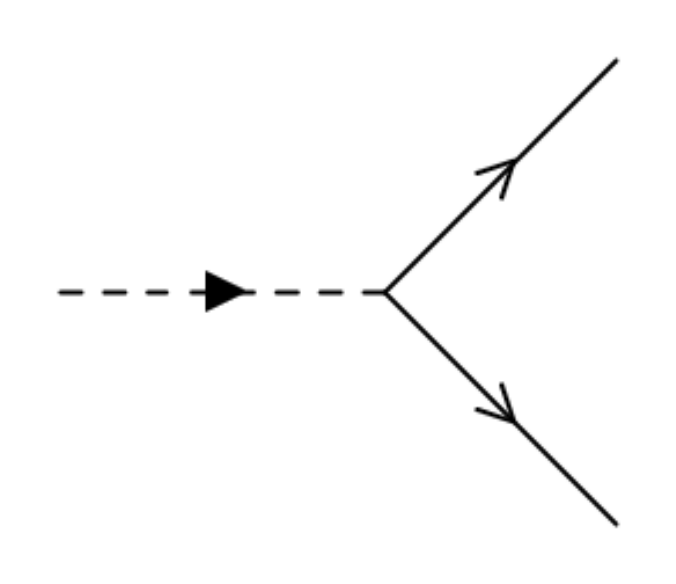}}
\caption{(a) The lowest order correction to the normal Green function. Straight dashed line corresponds to the static bosonic field, $\left< b \right>$. (b) Coupling between a fermion and the static bosonic field, here only the $c^\dagger c^\dagger$ interaction is shown.}
\label{Fig.main}
\end{figure}

As an illustration, the leading contribution to the fermionic self-energy due to the coupling is then given in Figure A\ref{fig_feymann_2},
\begin{equation}
    \Sigma(p,i\eta_m)=-\frac{1}{\beta}\sum_{\tilde{p}}\sum_{i\omega_n}|\Tilde{V}|^2 D(\tilde{p},i\omega_n)G^{(0)}(\tilde{p}-p,i(\omega_n-\eta_m)),
    \label{selfene}
\end{equation}
where $D(p,i\omega_n)$ denotes the finite-temperature time-ordered Green functions for the bosons and $G^{(0)}(p,i\eta_m)$ for the fermions.
For a representative bosonic Green function with a pole near $\epsilon_p$, Eq.\ref{selfene} is roughly
\begin{equation}
\label{SigEq}
    \Sigma(p,i\eta_m)=\sum_{\tilde{p}} |\Tilde{V}|^2\frac{1}{-i\eta_m-\xi_{\tilde{p}-p}+\epsilon_{\tilde{p}}}n_B(\epsilon_{\tilde{p}}),
\end{equation}
given fermionic Green functions with poles near $\xi_{p}$. It is straightforward to confirm that the integration over momentum gives
\begin{equation}
\begin{aligned}
     \Sigma(p,i\eta_m)&\approx|\Tilde{V}|^2 \frac{N}{-i\eta_m-\xi_{p_0-p} + \epsilon_{p_{0}}}\\
     &\approx |\Tilde{V}|^2 \frac{N}{-i\eta_m-\xi_{p_0-p}}
     \end{aligned}
     \label{equ_nonmf}
\end{equation}
corresponding to the diagram in Fig.A\ref{fig_feymann_3}, where $N$ denotes the number of bosons and we have made use of the large energy separation ($\xi_{p} \gg \epsilon_{p}$) discussed above. Altogether, this is equivalent to replacing the original coupling in Fig.A\ref{fig_feymann_1} by a static average field in Fig.A\ref{fig_feymann_4}, or equivalently replacing $H_{\text{coupling}}$ in Eq.\ref{SuppBogo} with

\begin{equation}
\label{AvgStatField}
    H_{\text{coupling}}\approx\sum_{p}\Tilde{V}\sqrt{N}c_{p}c_{p_0-p}+h.c.
\end{equation}

This example clearly demonstrates that, as long as the dynamics of the bosons are much slower than that of the fermions, one can approximate the effect on the fermions as coming from a static average with an approximate momentum $p_{0}$ as represented in Eq.\ref{AvgStatField}. 
Such simplification is even better justified when the physical broadening, as illustrated in Fig.2, is significant. In our case, such boradening occurs due to the lack of translational symmetry in the phase structure of the bosonic field, which couples all momenta in a line that spans the Brillouin zone.

\subsection{B. Bose Metal Effective Mean Field}
In Appendix A, we identified a simple average static field form to approximate the coupling to the bosons above given by Eq.(\ref{AvgStatField}) and illustrated in Fig.A\ref{fig_feymann_4}. Here we reproduce an explicit form of that average static field for a Bose metal as outlined in a previously published work\cite{hegg2021bose}. For completeness, we include the seemingly complicated formula in detail resulting simply from the mismatch between the boson and fermion lattices that bears little significance to the general features of this work.

The effective mean field is given by the form of the boson dispersion minimum. For the Bose metal\cite{hegg2021bose} this is simply
\begin{equation}
\label{E1}
<\tilde{b}^\dagger_{x+\frac{1}{2}\,,\,y\,,\,z}>=<\tilde{b}_{x+\frac{1}{2}\,,\,y\,,\,z}>=\sqrt{\frac{\delta}{4}}e^{\,i\,[\,\theta_y+\pi\,x\,]},
\end{equation}
for horizontal bonds and
\begin{equation}
\label{E2}
<\tilde{b}^\dagger_{x\,,\,y+\frac{1}{2}\,,\,z}>=<\tilde{b}_{x\,,\,y+\frac{1}{2}\,,\,z}>=\sqrt{\frac{\delta}{4}}e^{\,i\,[\,\theta_x+\pi\,y\,]},
\end{equation}
for vertical bonds, where $\delta$ is the doping fraction and $\tilde{b}^\dagger_{x+\frac{1}{2},y}$\Big($\tilde{b}^\dagger_{x,y+\frac{1}{2}}$\Big) creates a boson coherent along the $\hat{x}$($\hat{y}$) direction. Here $\theta_x$($\theta_y$) must be chosen at random from $0$ to $2\pi$, which reflects the lack of coherence due to geometric frustration in this Bose metal\cite{hegg2021bose}.

The convention used in the main text, $b_{i\,j}$, represents a boson between sites $i$ and $j$. These two representations are related via
\begin{equation}
b_{(x\,,\,y\,,\,z)\,(x\,,\,y+1\,,\,z)}=(-1)^y \tilde{b}_{x\,,\,y+\frac{1}{2}\,,\,z}
\end{equation}
\begin{equation}
b_{(x\,,\,y\,,\,z)\,(x+1\,,\,y\,,\,z)}=(-1)^x \tilde{b}_{x+\frac{1}{2}\,,\,y\,,\,z}, 
\end{equation}
where the sign change is due to fermionic commutation relations (see the supplementary material of \cite{jiang2019non} for details).
For our mean-field-like treatment, we find
\begin{equation}
\label{E4}
\begin{aligned}
H_{\text{coupling}}=\sum_{k_x k_y}&\Big(\sum_{q}V_y(q)c_{k_x,k_y}c_{-k_x,-k_y+q}\\
    &+\sum_{q}V_x(q)c_{k_x,k_y}c_{-k_x+q,-k_y}\Big)+ h.c.\\
\end{aligned}
\end{equation}
where
\begin{equation}
\label{Vs1}
V_y(q)=\frac{1}{N}\sqrt{\frac{\delta}{4}}V\sum_y e^{\,i\,[\,\theta_y+a\,q\,y\,]}
\end{equation}
\begin{equation}
\label{Vs2}
V_x(q)=\frac{1}{N}\sqrt{\frac{\delta}{4}}V\sum_{x}
e^{\,i\,[\,\theta_x+a\,q\,x\,]}.
\end{equation}

Note that although the static field denoted by Eqs.(\ref{Vs1},\ref{Vs2}) has a non-trivial form, it has no tunable parameters. Its phase structure simply represents the static structure of the dispersion minimum for the Bose metal\cite{hegg2021bose}, and its amplitude is fixed by the total carrier density.

\subsection{C. Underlying emergent Bose liquid from a microscopic consideration and its repeated success in the cuprates}

This manuscript aims to establish a robust and generic phenomenological connection between the imperfect coherence in the recently proposed Bose metal~\cite{hegg2021bose} and both the unusual non-Fermi liquid scattering rate and pseudogap formation when fermions couple to it. This connection is independent of any material specifics. Nonetheless, since these unusual phenomena are prominently observed in the unconventional superconducting cuprates, in this section we summarize the basic assumptions of the underlying emergent Bose liquid (EBL) model and how this lower-energy model incorporates realistic considerations for the cuprates obtained from widely studied high-energy models.

The Bose metal (employed in this study as a mean field) is obtained as one of the ground states of the EBL model~\cite{yildirim2011kinetics}. As the first assumption of the EBL model~\cite{yildirim2011kinetics,yildirim2015weak,jiang2019non,zeng2021transport,lang2022mottness,hegg2021bose}, the charge carriers are assumed to be mostly in a tightly bound two-body (bi-fermion) charge-2$e$ state involving the nearest neighboring atomic sites. This assumption is natural below some energy scale (say 150meV) in the presence of various high-energy physics previously proposed for the cuprates, for example, bi-polaronic correlation~\cite{alexandrov1994bipolarons,PhysRevX.13.011010}, two-dimensional short-range anti-ferromagnetic correlation~\cite{PhysRevLett.106.036401,PhysRevB.55.6504,altman2002plaquette} and/or cancellation of topological spin current~\cite{PhysRevLett.123.016601,zhao2022two}. In hole-doped cuprates, when using the half-filled Mott insulator with strong short-range antiferromagnetic correlation as the ``vacumm'' reference, this \textit{bond-centered} bosonic bi-fermion state would consist of doped holes of opposite spin residing in neighboring Cu sites, each as part of a Zhang-Rice singlet~\cite{zhang1988effective} with an intrinsic hole. For the square fermionic lattice of copper in the cuprates, the corresponding bond lattice of the boson containing the leading (non-pair breaking) kinetic process is therefore a checkerboard lattice. (A more detailed derivation is available in Ref.~\cite{yildirim2011kinetics}.)

As the second assumption of the EBL model, the underlying fermions are not allowed to reside on the same sites due to even higher-energy physics (for example the double-occupation constraint of the $t$-$J$ model resulting from the strong intra-atomic repulsion). This leads to the ``extended hardcore constraint'' of the EBL model that forbids occupation of the nearby sites around a boson by another boson.  For the above checkerboard bosonic lattice of the cuprates, six nearby sites are ``blocked'' by each boson.  (An important consequence of this unusual constraint is suppression of the superfluid stiffness in the overdoped region and ultimate destruction of superconductivity around $25\%$. See Ref.~\cite{lang2022mottness} for more details.)

Despite its simplicity, a series of previous studies based on this EBL model have successfully explained many unusual properties of the cuprates. This includes for example, the unusual momentum dependence of the superconducting gap~\cite{yildirim2011kinetics}, destruction of superconductivity and unusual properties near $5\%$ doping~\cite{yildirim2015weak}, anti-intuitive reduction of superfluid stiffness upon increasing carrier density and the overdoped critical doping~\cite{lang2022mottness}, the non-superfluid low-temperature pseudogap phase as a Bose metal ~\cite{hegg2021bose},  features in the exotic optical conductivity (mid-infrared and ubiquitous continuum) and transport properties (bad metal, strange metal, and weak insulating behavior)~\cite{zeng2021transport}, and proportionality between the low-temperature superconducting gap and the normal-state scattering rate~\cite{jiang2019non}. Most essentially, all the disparate semi-quantitative results above were obtained within the \textit{same} Hamiltonian under a \textit{fixed} set of parameters in the original work~\cite{yildirim2011kinetics}.

\bibliography{apssamp}

\begin{thebibliography}{82}%
\makeatletter
\providecommand \@ifxundefined [1]{%
 \@ifx{#1\undefined}
}%
\providecommand \@ifnum [1]{%
 \ifnum #1\expandafter \@firstoftwo
 \else \expandafter \@secondoftwo
 \fi
}%
\providecommand \@ifx [1]{%
 \ifx #1\expandafter \@firstoftwo
 \else \expandafter \@secondoftwo
 \fi
}%
\providecommand \natexlab [1]{#1}%
\providecommand \enquote  [1]{``#1''}%
\providecommand \bibnamefont  [1]{#1}%
\providecommand \bibfnamefont [1]{#1}%
\providecommand \citenamefont [1]{#1}%
\providecommand \href@noop [0]{\@secondoftwo}%
\providecommand \href [0]{\begingroup \@sanitize@url \@href}%
\providecommand \@href[1]{\@@startlink{#1}\@@href}%
\providecommand \@@href[1]{\endgroup#1\@@endlink}%
\providecommand \@sanitize@url [0]{\catcode `\\12\catcode `\$12\catcode
  `\&12\catcode `\#12\catcode `\^12\catcode `\_12\catcode `\%12\relax}%
\providecommand \@@startlink[1]{}%
\providecommand \@@endlink[0]{}%
\providecommand \url  [0]{\begingroup\@sanitize@url \@url }%
\providecommand \@url [1]{\endgroup\@href {#1}{\urlprefix }}%
\providecommand \urlprefix  [0]{URL }%
\providecommand \Eprint [0]{\href }%
\providecommand \doibase [0]{http://dx.doi.org/}%
\providecommand \selectlanguage [0]{\@gobble}%
\providecommand \bibinfo  [0]{\@secondoftwo}%
\providecommand \bibfield  [0]{\@secondoftwo}%
\providecommand \translation [1]{[#1]}%
\providecommand \BibitemOpen [0]{}%
\providecommand \bibitemStop [0]{}%
\providecommand \bibitemNoStop [0]{.\EOS\space}%
\providecommand \EOS [0]{\spacefactor3000\relax}%
\providecommand \BibitemShut  [1]{\csname bibitem#1\endcsname}%
\let\auto@bib@innerbib\@empty
%</preamble>
\bibitem [{\citenamefont {Hegg}\ \emph {et~al.}(2021)\citenamefont {Hegg},
  \citenamefont {Hou},\ and\ \citenamefont {Ku}}]{hegg2021bose}%
  \BibitemOpen
  \bibfield  {author} {\bibinfo {author} {\bibfnamefont {A.}~\bibnamefont
  {Hegg}}, \bibinfo {author} {\bibfnamefont {J.}~\bibnamefont {Hou}}, \ and\
  \bibinfo {author} {\bibfnamefont {W.}~\bibnamefont {Ku}},\ }\href {\doibase
  10.1073/pnas.2100545118} {\bibfield  {journal} {\bibinfo  {journal}
  {Proceedings of the National Academy of Sciences}\ }\textbf {\bibinfo
  {volume} {118}} (\bibinfo {year} {2021}),\
  10.1073/pnas.2100545118}\BibitemShut {NoStop}%
\bibitem [{\citenamefont {Dagotto}(1994)}]{dagotto1994correlated}%
  \BibitemOpen
  \bibfield  {author} {\bibinfo {author} {\bibfnamefont {E.}~\bibnamefont
  {Dagotto}},\ }\href@noop {} {\bibfield  {journal} {\bibinfo  {journal}
  {Reviews of Modern Physics}\ }\textbf {\bibinfo {volume} {66}},\ \bibinfo
  {pages} {763} (\bibinfo {year} {1994})}\BibitemShut {NoStop}%
\bibitem [{\citenamefont {Damascelli}\ \emph {et~al.}(2003)\citenamefont
  {Damascelli}, \citenamefont {Hussain},\ and\ \citenamefont
  {Shen}}]{damascelli2003angle}%
  \BibitemOpen
  \bibfield  {author} {\bibinfo {author} {\bibfnamefont {A.}~\bibnamefont
  {Damascelli}}, \bibinfo {author} {\bibfnamefont {Z.}~\bibnamefont {Hussain}},
  \ and\ \bibinfo {author} {\bibfnamefont {Z.-X.}\ \bibnamefont {Shen}},\
  }\href@noop {} {\bibfield  {journal} {\bibinfo  {journal} {Reviews of modern
  physics}\ }\textbf {\bibinfo {volume} {75}},\ \bibinfo {pages} {473}
  (\bibinfo {year} {2003})}\BibitemShut {NoStop}%
\bibitem [{\citenamefont {Lee}(2014)}]{lee2014amperean}%
  \BibitemOpen
  \bibfield  {author} {\bibinfo {author} {\bibfnamefont {P.~A.}\ \bibnamefont
  {Lee}},\ }\href@noop {} {\bibfield  {journal} {\bibinfo  {journal} {Physical
  Review X}\ }\textbf {\bibinfo {volume} {4}},\ \bibinfo {pages} {031017}
  (\bibinfo {year} {2014})}\BibitemShut {NoStop}%
\bibitem [{\citenamefont {Battisti}\ \emph {et~al.}(2017)\citenamefont
  {Battisti}, \citenamefont {Bastiaans}, \citenamefont {Fedoseev},
  \citenamefont {De~La~Torre}, \citenamefont {Iliopoulos}, \citenamefont
  {Tamai}, \citenamefont {Hunter}, \citenamefont {Perry}, \citenamefont
  {Zaanen}, \citenamefont {Baumberger} \emph
  {et~al.}}]{battisti2017universality}%
  \BibitemOpen
  \bibfield  {author} {\bibinfo {author} {\bibfnamefont {I.}~\bibnamefont
  {Battisti}}, \bibinfo {author} {\bibfnamefont {K.~M.}\ \bibnamefont
  {Bastiaans}}, \bibinfo {author} {\bibfnamefont {V.}~\bibnamefont {Fedoseev}},
  \bibinfo {author} {\bibfnamefont {A.}~\bibnamefont {De~La~Torre}}, \bibinfo
  {author} {\bibfnamefont {N.}~\bibnamefont {Iliopoulos}}, \bibinfo {author}
  {\bibfnamefont {A.}~\bibnamefont {Tamai}}, \bibinfo {author} {\bibfnamefont
  {E.~C.}\ \bibnamefont {Hunter}}, \bibinfo {author} {\bibfnamefont {R.~S.}\
  \bibnamefont {Perry}}, \bibinfo {author} {\bibfnamefont {J.}~\bibnamefont
  {Zaanen}}, \bibinfo {author} {\bibfnamefont {F.}~\bibnamefont {Baumberger}},
  \emph {et~al.},\ }\href@noop {} {\bibfield  {journal} {\bibinfo  {journal}
  {Nature Physics}\ }\textbf {\bibinfo {volume} {13}},\ \bibinfo {pages} {21}
  (\bibinfo {year} {2017})}\BibitemShut {NoStop}%
\bibitem [{\citenamefont {Abrikosov}\ \emph {et~al.}(2012)\citenamefont
  {Abrikosov}, \citenamefont {Gorkov},\ and\ \citenamefont
  {Dzyaloshinski}}]{abrikosov}%
  \BibitemOpen
  \bibfield  {author} {\bibinfo {author} {\bibfnamefont {A.~A.}\ \bibnamefont
  {Abrikosov}}, \bibinfo {author} {\bibfnamefont {L.~P.}\ \bibnamefont
  {Gorkov}}, \ and\ \bibinfo {author} {\bibfnamefont {I.~E.}\ \bibnamefont
  {Dzyaloshinski}},\ }\href@noop {} {\emph {\bibinfo {title} {Methods of
  quantum field theory in statistical physics}}}\ (\bibinfo  {publisher}
  {Courier Corporation},\ \bibinfo {year} {2012})\BibitemShut {NoStop}%
\bibitem [{\citenamefont {Valla}\ \emph {et~al.}(1999)\citenamefont {Valla},
  \citenamefont {Fedorov}, \citenamefont {Johnson}, \citenamefont {Wells},
  \citenamefont {Hulbert}, \citenamefont {Li}, \citenamefont {Gu},\ and\
  \citenamefont {Koshizuka}}]{valla1999evidence}%
  \BibitemOpen
  \bibfield  {author} {\bibinfo {author} {\bibfnamefont {T.}~\bibnamefont
  {Valla}}, \bibinfo {author} {\bibfnamefont {A.}~\bibnamefont {Fedorov}},
  \bibinfo {author} {\bibfnamefont {P.}~\bibnamefont {Johnson}}, \bibinfo
  {author} {\bibfnamefont {B.}~\bibnamefont {Wells}}, \bibinfo {author}
  {\bibfnamefont {S.}~\bibnamefont {Hulbert}}, \bibinfo {author} {\bibfnamefont
  {Q.}~\bibnamefont {Li}}, \bibinfo {author} {\bibfnamefont {G.}~\bibnamefont
  {Gu}}, \ and\ \bibinfo {author} {\bibfnamefont {N.}~\bibnamefont
  {Koshizuka}},\ }\href@noop {} {\bibfield  {journal} {\bibinfo  {journal}
  {Science}\ }\textbf {\bibinfo {volume} {285}},\ \bibinfo {pages} {2110}
  (\bibinfo {year} {1999})}\BibitemShut {NoStop}%
\bibitem [{\citenamefont {Abdel-Jawad}\ \emph {et~al.}(2006)\citenamefont
  {Abdel-Jawad}, \citenamefont {Kennett}, \citenamefont {Balicas},
  \citenamefont {Carrington}, \citenamefont {Mackenzie}, \citenamefont
  {McKenzie},\ and\ \citenamefont {Hussey}}]{abdel2006anisotropic}%
  \BibitemOpen
  \bibfield  {author} {\bibinfo {author} {\bibfnamefont {M.}~\bibnamefont
  {Abdel-Jawad}}, \bibinfo {author} {\bibfnamefont {M.}~\bibnamefont
  {Kennett}}, \bibinfo {author} {\bibfnamefont {L.}~\bibnamefont {Balicas}},
  \bibinfo {author} {\bibfnamefont {A.}~\bibnamefont {Carrington}}, \bibinfo
  {author} {\bibfnamefont {A.}~\bibnamefont {Mackenzie}}, \bibinfo {author}
  {\bibfnamefont {R.}~\bibnamefont {McKenzie}}, \ and\ \bibinfo {author}
  {\bibfnamefont {N.}~\bibnamefont {Hussey}},\ }\href@noop {} {\bibfield
  {journal} {\bibinfo  {journal} {Nature Physics}\ }\textbf {\bibinfo {volume}
  {2}},\ \bibinfo {pages} {821} (\bibinfo {year} {2006})}\BibitemShut {NoStop}%
\bibitem [{\citenamefont {Johnston}\ \emph {et~al.}(2012)\citenamefont
  {Johnston}, \citenamefont {Vishik}, \citenamefont {Lee}, \citenamefont
  {Schmitt}, \citenamefont {Uchida}, \citenamefont {Fujita}, \citenamefont
  {Ishida}, \citenamefont {Nagaosa}, \citenamefont {Shen},\ and\ \citenamefont
  {Devereaux}}]{johnston2012evidence}%
  \BibitemOpen
  \bibfield  {author} {\bibinfo {author} {\bibfnamefont {S.}~\bibnamefont
  {Johnston}}, \bibinfo {author} {\bibfnamefont {I.}~\bibnamefont {Vishik}},
  \bibinfo {author} {\bibfnamefont {W.}~\bibnamefont {Lee}}, \bibinfo {author}
  {\bibfnamefont {F.}~\bibnamefont {Schmitt}}, \bibinfo {author} {\bibfnamefont
  {S.}~\bibnamefont {Uchida}}, \bibinfo {author} {\bibfnamefont
  {K.}~\bibnamefont {Fujita}}, \bibinfo {author} {\bibfnamefont
  {S.}~\bibnamefont {Ishida}}, \bibinfo {author} {\bibfnamefont
  {N.}~\bibnamefont {Nagaosa}}, \bibinfo {author} {\bibfnamefont
  {Z.}~\bibnamefont {Shen}}, \ and\ \bibinfo {author} {\bibfnamefont
  {T.}~\bibnamefont {Devereaux}},\ }\href@noop {} {\bibfield  {journal}
  {\bibinfo  {journal} {Physical review letters}\ }\textbf {\bibinfo {volume}
  {108}},\ \bibinfo {pages} {166404} (\bibinfo {year} {2012})}\BibitemShut
  {NoStop}%
\bibitem [{\citenamefont {Kaminski}\ \emph {et~al.}(2005)\citenamefont
  {Kaminski}, \citenamefont {Fretwell}, \citenamefont {Norman}, \citenamefont
  {Randeria}, \citenamefont {Rosenkranz}, \citenamefont {Chatterjee},
  \citenamefont {Campuzano}, \citenamefont {Mesot}, \citenamefont {Sato},
  \citenamefont {Takahashi} \emph {et~al.}}]{kaminski2005momentum}%
  \BibitemOpen
  \bibfield  {author} {\bibinfo {author} {\bibfnamefont {A.}~\bibnamefont
  {Kaminski}}, \bibinfo {author} {\bibfnamefont {H.}~\bibnamefont {Fretwell}},
  \bibinfo {author} {\bibfnamefont {M.~R.}\ \bibnamefont {Norman}}, \bibinfo
  {author} {\bibfnamefont {M.}~\bibnamefont {Randeria}}, \bibinfo {author}
  {\bibfnamefont {S.}~\bibnamefont {Rosenkranz}}, \bibinfo {author}
  {\bibfnamefont {U.}~\bibnamefont {Chatterjee}}, \bibinfo {author}
  {\bibfnamefont {J.}~\bibnamefont {Campuzano}}, \bibinfo {author}
  {\bibfnamefont {J.}~\bibnamefont {Mesot}}, \bibinfo {author} {\bibfnamefont
  {T.}~\bibnamefont {Sato}}, \bibinfo {author} {\bibfnamefont {T.}~\bibnamefont
  {Takahashi}},  \emph {et~al.},\ }\href@noop {} {\bibfield  {journal}
  {\bibinfo  {journal} {Physical review B}\ }\textbf {\bibinfo {volume} {71}},\
  \bibinfo {pages} {014517} (\bibinfo {year} {2005})}\BibitemShut {NoStop}%
\bibitem [{\citenamefont {Bruin}\ \emph {et~al.}(2013)\citenamefont {Bruin},
  \citenamefont {Sakai}, \citenamefont {Perry},\ and\ \citenamefont
  {Mackenzie}}]{bruin2013similarity}%
  \BibitemOpen
  \bibfield  {author} {\bibinfo {author} {\bibfnamefont {J.}~\bibnamefont
  {Bruin}}, \bibinfo {author} {\bibfnamefont {H.}~\bibnamefont {Sakai}},
  \bibinfo {author} {\bibfnamefont {R.}~\bibnamefont {Perry}}, \ and\ \bibinfo
  {author} {\bibfnamefont {A.}~\bibnamefont {Mackenzie}},\ }\href@noop {}
  {\bibfield  {journal} {\bibinfo  {journal} {Science}\ }\textbf {\bibinfo
  {volume} {339}},\ \bibinfo {pages} {804} (\bibinfo {year}
  {2013})}\BibitemShut {NoStop}%
\bibitem [{\citenamefont {Fink}\ \emph {et~al.}(2015)\citenamefont {Fink},
  \citenamefont {Charnukha}, \citenamefont {Rienks}, \citenamefont {Liu},
  \citenamefont {Thirupathaiah}, \citenamefont {Avigo}, \citenamefont {Roth},
  \citenamefont {Jeevan}, \citenamefont {Gegenwart}, \citenamefont {Roslova}
  \emph {et~al.}}]{fink2015non}%
  \BibitemOpen
  \bibfield  {author} {\bibinfo {author} {\bibfnamefont {J.}~\bibnamefont
  {Fink}}, \bibinfo {author} {\bibfnamefont {A.}~\bibnamefont {Charnukha}},
  \bibinfo {author} {\bibfnamefont {E.}~\bibnamefont {Rienks}}, \bibinfo
  {author} {\bibfnamefont {Z.}~\bibnamefont {Liu}}, \bibinfo {author}
  {\bibfnamefont {S.}~\bibnamefont {Thirupathaiah}}, \bibinfo {author}
  {\bibfnamefont {I.}~\bibnamefont {Avigo}}, \bibinfo {author} {\bibfnamefont
  {F.}~\bibnamefont {Roth}}, \bibinfo {author} {\bibfnamefont {H.~S.}\
  \bibnamefont {Jeevan}}, \bibinfo {author} {\bibfnamefont {P.}~\bibnamefont
  {Gegenwart}}, \bibinfo {author} {\bibfnamefont {M.}~\bibnamefont {Roslova}},
  \emph {et~al.},\ }\href@noop {} {\bibfield  {journal} {\bibinfo  {journal}
  {Physical Review B}\ }\textbf {\bibinfo {volume} {92}},\ \bibinfo {pages}
  {201106} (\bibinfo {year} {2015})}\BibitemShut {NoStop}%
\bibitem [{\citenamefont {Dai}\ \emph {et~al.}(2015)\citenamefont {Dai},
  \citenamefont {Miao}, \citenamefont {Xing}, \citenamefont {Wang},
  \citenamefont {Wang}, \citenamefont {Xiao}, \citenamefont {Qian},
  \citenamefont {Richard}, \citenamefont {Qiu}, \citenamefont {Yu} \emph
  {et~al.}}]{dai2015spin}%
  \BibitemOpen
  \bibfield  {author} {\bibinfo {author} {\bibfnamefont {Y.}~\bibnamefont
  {Dai}}, \bibinfo {author} {\bibfnamefont {H.}~\bibnamefont {Miao}}, \bibinfo
  {author} {\bibfnamefont {L.}~\bibnamefont {Xing}}, \bibinfo {author}
  {\bibfnamefont {X.}~\bibnamefont {Wang}}, \bibinfo {author} {\bibfnamefont
  {P.}~\bibnamefont {Wang}}, \bibinfo {author} {\bibfnamefont {H.}~\bibnamefont
  {Xiao}}, \bibinfo {author} {\bibfnamefont {T.}~\bibnamefont {Qian}}, \bibinfo
  {author} {\bibfnamefont {P.}~\bibnamefont {Richard}}, \bibinfo {author}
  {\bibfnamefont {X.}~\bibnamefont {Qiu}}, \bibinfo {author} {\bibfnamefont
  {W.}~\bibnamefont {Yu}},  \emph {et~al.},\ }\href@noop {} {\bibfield
  {journal} {\bibinfo  {journal} {Physical Review X}\ }\textbf {\bibinfo
  {volume} {5}},\ \bibinfo {pages} {031035} (\bibinfo {year}
  {2015})}\BibitemShut {NoStop}%
\bibitem [{\citenamefont {Umemoto}\ \emph {et~al.}(2019)\citenamefont
  {Umemoto}, \citenamefont {Sugawara}, \citenamefont {Nakata}, \citenamefont
  {Takahashi},\ and\ \citenamefont {Sato}}]{umemoto2019pseudogap}%
  \BibitemOpen
  \bibfield  {author} {\bibinfo {author} {\bibfnamefont {Y.}~\bibnamefont
  {Umemoto}}, \bibinfo {author} {\bibfnamefont {K.}~\bibnamefont {Sugawara}},
  \bibinfo {author} {\bibfnamefont {Y.}~\bibnamefont {Nakata}}, \bibinfo
  {author} {\bibfnamefont {T.}~\bibnamefont {Takahashi}}, \ and\ \bibinfo
  {author} {\bibfnamefont {T.}~\bibnamefont {Sato}},\ }\href@noop {} {\bibfield
   {journal} {\bibinfo  {journal} {Nano Research}\ }\textbf {\bibinfo {volume}
  {12}},\ \bibinfo {pages} {165} (\bibinfo {year} {2019})}\BibitemShut
  {NoStop}%
\bibitem [{\citenamefont {L{\"o}hneysen}\ \emph {et~al.}(1994)\citenamefont
  {L{\"o}hneysen}, \citenamefont {Pietrus}, \citenamefont {Portisch},
  \citenamefont {Schlager}, \citenamefont {Schr{\"o}der}, \citenamefont
  {Sieck},\ and\ \citenamefont {Trappmann}}]{lohneysen1994non}%
  \BibitemOpen
  \bibfield  {author} {\bibinfo {author} {\bibfnamefont {H.~v.}\ \bibnamefont
  {L{\"o}hneysen}}, \bibinfo {author} {\bibfnamefont {T.}~\bibnamefont
  {Pietrus}}, \bibinfo {author} {\bibfnamefont {G.}~\bibnamefont {Portisch}},
  \bibinfo {author} {\bibfnamefont {H.}~\bibnamefont {Schlager}}, \bibinfo
  {author} {\bibfnamefont {A.}~\bibnamefont {Schr{\"o}der}}, \bibinfo {author}
  {\bibfnamefont {M.}~\bibnamefont {Sieck}}, \ and\ \bibinfo {author}
  {\bibfnamefont {T.}~\bibnamefont {Trappmann}},\ }\href@noop {} {\bibfield
  {journal} {\bibinfo  {journal} {Physical review letters}\ }\textbf {\bibinfo
  {volume} {72}},\ \bibinfo {pages} {3262} (\bibinfo {year}
  {1994})}\BibitemShut {NoStop}%
\bibitem [{\citenamefont {Paglione}\ \emph {et~al.}(2007)\citenamefont
  {Paglione}, \citenamefont {Sayles}, \citenamefont {Ho}, \citenamefont
  {Jeffries},\ and\ \citenamefont {Maple}}]{paglione2007incoherent}%
  \BibitemOpen
  \bibfield  {author} {\bibinfo {author} {\bibfnamefont {J.}~\bibnamefont
  {Paglione}}, \bibinfo {author} {\bibfnamefont {T.}~\bibnamefont {Sayles}},
  \bibinfo {author} {\bibfnamefont {P.-C.}\ \bibnamefont {Ho}}, \bibinfo
  {author} {\bibfnamefont {J.}~\bibnamefont {Jeffries}}, \ and\ \bibinfo
  {author} {\bibfnamefont {M.}~\bibnamefont {Maple}},\ }\href@noop {}
  {\bibfield  {journal} {\bibinfo  {journal} {Nature Physics}\ }\textbf
  {\bibinfo {volume} {3}},\ \bibinfo {pages} {703} (\bibinfo {year}
  {2007})}\BibitemShut {NoStop}%
\bibitem [{\citenamefont {Tao}\ \emph {et~al.}(1997)\citenamefont {Tao},
  \citenamefont {Lu},\ and\ \citenamefont {Wolf}}]{tao1997observation}%
  \BibitemOpen
  \bibfield  {author} {\bibinfo {author} {\bibfnamefont {H.}~\bibnamefont
  {Tao}}, \bibinfo {author} {\bibfnamefont {F.}~\bibnamefont {Lu}}, \ and\
  \bibinfo {author} {\bibfnamefont {E.}~\bibnamefont {Wolf}},\ }\href@noop {}
  {\bibfield  {journal} {\bibinfo  {journal} {Physica C: Superconductivity}\
  }\textbf {\bibinfo {volume} {282}},\ \bibinfo {pages} {1507} (\bibinfo {year}
  {1997})}\BibitemShut {NoStop}%
\bibitem [{\citenamefont {Renner}\ \emph {et~al.}(1998)\citenamefont {Renner},
  \citenamefont {Revaz}, \citenamefont {Genoud}, \citenamefont {Kadowaki},\
  and\ \citenamefont {Fischer}}]{renner1998pseudogap}%
  \BibitemOpen
  \bibfield  {author} {\bibinfo {author} {\bibfnamefont {C.}~\bibnamefont
  {Renner}}, \bibinfo {author} {\bibfnamefont {B.}~\bibnamefont {Revaz}},
  \bibinfo {author} {\bibfnamefont {J.-Y.}\ \bibnamefont {Genoud}}, \bibinfo
  {author} {\bibfnamefont {K.}~\bibnamefont {Kadowaki}}, \ and\ \bibinfo
  {author} {\bibfnamefont {{\O}.}~\bibnamefont {Fischer}},\ }\href@noop {}
  {\bibfield  {journal} {\bibinfo  {journal} {Physical Review Letters}\
  }\textbf {\bibinfo {volume} {80}},\ \bibinfo {pages} {149} (\bibinfo {year}
  {1998})}\BibitemShut {NoStop}%
\bibitem [{\citenamefont {Timusk}\ and\ \citenamefont
  {Statt}(1999)}]{timusk1999pseudogap}%
  \BibitemOpen
  \bibfield  {author} {\bibinfo {author} {\bibfnamefont {T.}~\bibnamefont
  {Timusk}}\ and\ \bibinfo {author} {\bibfnamefont {B.}~\bibnamefont {Statt}},\
  }\href@noop {} {\bibfield  {journal} {\bibinfo  {journal} {Reports on
  Progress in Physics}\ }\textbf {\bibinfo {volume} {62}},\ \bibinfo {pages}
  {61} (\bibinfo {year} {1999})}\BibitemShut {NoStop}%
\bibitem [{\citenamefont {Lang}\ \emph {et~al.}(2002)\citenamefont {Lang},
  \citenamefont {Madhavan}, \citenamefont {Hoffman}, \citenamefont {Hudson},
  \citenamefont {Eisaki}, \citenamefont {Uchida},\ and\ \citenamefont
  {Davis}}]{lang2002imaging}%
  \BibitemOpen
  \bibfield  {author} {\bibinfo {author} {\bibfnamefont {K.}~\bibnamefont
  {Lang}}, \bibinfo {author} {\bibfnamefont {V.}~\bibnamefont {Madhavan}},
  \bibinfo {author} {\bibfnamefont {J.}~\bibnamefont {Hoffman}}, \bibinfo
  {author} {\bibfnamefont {E.~W.}\ \bibnamefont {Hudson}}, \bibinfo {author}
  {\bibfnamefont {H.}~\bibnamefont {Eisaki}}, \bibinfo {author} {\bibfnamefont
  {S.}~\bibnamefont {Uchida}}, \ and\ \bibinfo {author} {\bibfnamefont
  {J.}~\bibnamefont {Davis}},\ }\href@noop {} {\bibfield  {journal} {\bibinfo
  {journal} {Nature}\ }\textbf {\bibinfo {volume} {415}},\ \bibinfo {pages}
  {412} (\bibinfo {year} {2002})}\BibitemShut {NoStop}%
\bibitem [{\citenamefont {Yang}\ \emph {et~al.}(2008)\citenamefont {Yang},
  \citenamefont {Rameau}, \citenamefont {Johnson}, \citenamefont {Valla},
  \citenamefont {Tsvelik},\ and\ \citenamefont {Gu}}]{yang2008emergence}%
  \BibitemOpen
  \bibfield  {author} {\bibinfo {author} {\bibfnamefont {H.-B.}\ \bibnamefont
  {Yang}}, \bibinfo {author} {\bibfnamefont {J.}~\bibnamefont {Rameau}},
  \bibinfo {author} {\bibfnamefont {P.}~\bibnamefont {Johnson}}, \bibinfo
  {author} {\bibfnamefont {T.}~\bibnamefont {Valla}}, \bibinfo {author}
  {\bibfnamefont {A.}~\bibnamefont {Tsvelik}}, \ and\ \bibinfo {author}
  {\bibfnamefont {G.}~\bibnamefont {Gu}},\ }\href@noop {} {\bibfield  {journal}
  {\bibinfo  {journal} {Nature}\ }\textbf {\bibinfo {volume} {456}},\ \bibinfo
  {pages} {77} (\bibinfo {year} {2008})}\BibitemShut {NoStop}%
\bibitem [{\citenamefont {Hashimoto}\ \emph {et~al.}(2010)\citenamefont
  {Hashimoto}, \citenamefont {He}, \citenamefont {Tanaka}, \citenamefont
  {Testaud}, \citenamefont {Meevasana}, \citenamefont {Moore}, \citenamefont
  {Lu}, \citenamefont {Yao}, \citenamefont {Yoshida}, \citenamefont {Eisaki}
  \emph {et~al.}}]{particlehole}%
  \BibitemOpen
  \bibfield  {author} {\bibinfo {author} {\bibfnamefont {M.}~\bibnamefont
  {Hashimoto}}, \bibinfo {author} {\bibfnamefont {R.-H.}\ \bibnamefont {He}},
  \bibinfo {author} {\bibfnamefont {K.}~\bibnamefont {Tanaka}}, \bibinfo
  {author} {\bibfnamefont {J.-P.}\ \bibnamefont {Testaud}}, \bibinfo {author}
  {\bibfnamefont {W.}~\bibnamefont {Meevasana}}, \bibinfo {author}
  {\bibfnamefont {R.~G.}\ \bibnamefont {Moore}}, \bibinfo {author}
  {\bibfnamefont {D.}~\bibnamefont {Lu}}, \bibinfo {author} {\bibfnamefont
  {H.}~\bibnamefont {Yao}}, \bibinfo {author} {\bibfnamefont {Y.}~\bibnamefont
  {Yoshida}}, \bibinfo {author} {\bibfnamefont {H.}~\bibnamefont {Eisaki}},
  \emph {et~al.},\ }\href@noop {} {\bibfield  {journal} {\bibinfo  {journal}
  {Nature Physics}\ }\textbf {\bibinfo {volume} {6}},\ \bibinfo {pages} {414}
  (\bibinfo {year} {2010})}\BibitemShut {NoStop}%
\bibitem [{\citenamefont {Kondo}\ \emph {et~al.}(2011)\citenamefont {Kondo},
  \citenamefont {Hamaya}, \citenamefont {Palczewski}, \citenamefont {Takeuchi},
  \citenamefont {Wen}, \citenamefont {Xu}, \citenamefont {Gu}, \citenamefont
  {Schmalian},\ and\ \citenamefont {Kaminski}}]{pseudogap_t_depen}%
  \BibitemOpen
  \bibfield  {author} {\bibinfo {author} {\bibfnamefont {T.}~\bibnamefont
  {Kondo}}, \bibinfo {author} {\bibfnamefont {Y.}~\bibnamefont {Hamaya}},
  \bibinfo {author} {\bibfnamefont {A.~D.}\ \bibnamefont {Palczewski}},
  \bibinfo {author} {\bibfnamefont {T.}~\bibnamefont {Takeuchi}}, \bibinfo
  {author} {\bibfnamefont {J.~S.}\ \bibnamefont {Wen}}, \bibinfo {author}
  {\bibfnamefont {Z.~J.}\ \bibnamefont {Xu}}, \bibinfo {author} {\bibfnamefont
  {G.}~\bibnamefont {Gu}}, \bibinfo {author} {\bibfnamefont {J.}~\bibnamefont
  {Schmalian}}, \ and\ \bibinfo {author} {\bibfnamefont {A.}~\bibnamefont
  {Kaminski}},\ }\href {\doibase 10.1038/nphys1851} {\bibfield  {journal}
  {\bibinfo  {journal} {Nature Physics}\ }\textbf {\bibinfo {volume} {7}},\
  \bibinfo {pages} {21} (\bibinfo {year} {2011})}\BibitemShut {NoStop}%
\bibitem [{\citenamefont {He}\ \emph {et~al.}(2011)\citenamefont {He},
  \citenamefont {Hashimoto}, \citenamefont {Karapetyan}, \citenamefont
  {Koralek}, \citenamefont {Hinton}, \citenamefont {Testaud}, \citenamefont
  {Nathan}, \citenamefont {Yoshida}, \citenamefont {Yao}, \citenamefont
  {Tanaka} \emph {et~al.}}]{he2011single}%
  \BibitemOpen
  \bibfield  {author} {\bibinfo {author} {\bibfnamefont {R.-H.}\ \bibnamefont
  {He}}, \bibinfo {author} {\bibfnamefont {M.}~\bibnamefont {Hashimoto}},
  \bibinfo {author} {\bibfnamefont {H.}~\bibnamefont {Karapetyan}}, \bibinfo
  {author} {\bibfnamefont {J.}~\bibnamefont {Koralek}}, \bibinfo {author}
  {\bibfnamefont {J.}~\bibnamefont {Hinton}}, \bibinfo {author} {\bibfnamefont
  {J.}~\bibnamefont {Testaud}}, \bibinfo {author} {\bibfnamefont
  {V.}~\bibnamefont {Nathan}}, \bibinfo {author} {\bibfnamefont
  {Y.}~\bibnamefont {Yoshida}}, \bibinfo {author} {\bibfnamefont
  {H.}~\bibnamefont {Yao}}, \bibinfo {author} {\bibfnamefont {K.}~\bibnamefont
  {Tanaka}},  \emph {et~al.},\ }\href@noop {} {\bibfield  {journal} {\bibinfo
  {journal} {Science}\ }\textbf {\bibinfo {volume} {331}},\ \bibinfo {pages}
  {1579} (\bibinfo {year} {2011})}\BibitemShut {NoStop}%
\bibitem [{\citenamefont {Vishik}\ \emph {et~al.}(2010)\citenamefont {Vishik},
  \citenamefont {Lee}, \citenamefont {He}, \citenamefont {Hashimoto},
  \citenamefont {Hussain}, \citenamefont {Devereaux},\ and\ \citenamefont
  {Shen}}]{arpes_cuprate}%
  \BibitemOpen
  \bibfield  {author} {\bibinfo {author} {\bibfnamefont {I.~M.}\ \bibnamefont
  {Vishik}}, \bibinfo {author} {\bibfnamefont {W.~S.}\ \bibnamefont {Lee}},
  \bibinfo {author} {\bibfnamefont {R.-H.}\ \bibnamefont {He}}, \bibinfo
  {author} {\bibfnamefont {M.}~\bibnamefont {Hashimoto}}, \bibinfo {author}
  {\bibfnamefont {Z.}~\bibnamefont {Hussain}}, \bibinfo {author} {\bibfnamefont
  {T.~P.}\ \bibnamefont {Devereaux}}, \ and\ \bibinfo {author} {\bibfnamefont
  {Z.-X.}\ \bibnamefont {Shen}},\ }\href {\doibase
  10.1088/1367-2630/12/10/105008} {\bibfield  {journal} {\bibinfo  {journal}
  {New Journal of Physics}\ }\textbf {\bibinfo {volume} {12}},\ \bibinfo
  {pages} {105008} (\bibinfo {year} {2010})}\BibitemShut {NoStop}%
\bibitem [{\citenamefont {Yoshida}\ \emph {et~al.}(2012)\citenamefont
  {Yoshida}, \citenamefont {Hashimoto}, \citenamefont {M.~Vishik},
  \citenamefont {Shen},\ and\ \citenamefont {Fujimori}}]{arpes_cuprate_2}%
  \BibitemOpen
  \bibfield  {author} {\bibinfo {author} {\bibfnamefont {T.}~\bibnamefont
  {Yoshida}}, \bibinfo {author} {\bibfnamefont {M.}~\bibnamefont {Hashimoto}},
  \bibinfo {author} {\bibfnamefont {I.}~\bibnamefont {M.~Vishik}}, \bibinfo
  {author} {\bibfnamefont {Z.-X.}\ \bibnamefont {Shen}}, \ and\ \bibinfo
  {author} {\bibfnamefont {A.}~\bibnamefont {Fujimori}},\ }\href {\doibase
  10.1143/JPSJ.81.011006} {\bibfield  {journal} {\bibinfo  {journal} {Journal
  of the Physical Society of Japan}\ }\textbf {\bibinfo {volume} {81}},\
  \bibinfo {pages} {011006} (\bibinfo {year} {2012})}\BibitemShut {NoStop}%
\bibitem [{\citenamefont {Uchida}\ \emph {et~al.}(2011)\citenamefont {Uchida},
  \citenamefont {Ishizaka}, \citenamefont {Hansmann}, \citenamefont {Kaneko},
  \citenamefont {Ishida}, \citenamefont {Yang}, \citenamefont {Kumai},
  \citenamefont {Toschi}, \citenamefont {Onose}, \citenamefont {Arita},
  \citenamefont {Held}, \citenamefont {Andersen}, \citenamefont {Shin},\ and\
  \citenamefont {Tokura}}]{nickelate}%
  \BibitemOpen
  \bibfield  {author} {\bibinfo {author} {\bibfnamefont {M.}~\bibnamefont
  {Uchida}}, \bibinfo {author} {\bibfnamefont {K.}~\bibnamefont {Ishizaka}},
  \bibinfo {author} {\bibfnamefont {P.}~\bibnamefont {Hansmann}}, \bibinfo
  {author} {\bibfnamefont {Y.}~\bibnamefont {Kaneko}}, \bibinfo {author}
  {\bibfnamefont {Y.}~\bibnamefont {Ishida}}, \bibinfo {author} {\bibfnamefont
  {X.}~\bibnamefont {Yang}}, \bibinfo {author} {\bibfnamefont {R.}~\bibnamefont
  {Kumai}}, \bibinfo {author} {\bibfnamefont {A.}~\bibnamefont {Toschi}},
  \bibinfo {author} {\bibfnamefont {Y.}~\bibnamefont {Onose}}, \bibinfo
  {author} {\bibfnamefont {R.}~\bibnamefont {Arita}}, \bibinfo {author}
  {\bibfnamefont {K.}~\bibnamefont {Held}}, \bibinfo {author} {\bibfnamefont
  {O.~K.}\ \bibnamefont {Andersen}}, \bibinfo {author} {\bibfnamefont
  {S.}~\bibnamefont {Shin}}, \ and\ \bibinfo {author} {\bibfnamefont
  {Y.}~\bibnamefont {Tokura}},\ }\href {\doibase
  10.1103/PhysRevLett.106.027001} {\bibfield  {journal} {\bibinfo  {journal}
  {Phys. Rev. Lett.}\ }\textbf {\bibinfo {volume} {106}},\ \bibinfo {pages}
  {027001} (\bibinfo {year} {2011})}\BibitemShut {NoStop}%
\bibitem [{\citenamefont {Kim}\ \emph {et~al.}(2014)\citenamefont {Kim},
  \citenamefont {Krupin}, \citenamefont {Denlinger}, \citenamefont {Bostwick},
  \citenamefont {Rotenberg}, \citenamefont {Zhao}, \citenamefont {Mitchell},
  \citenamefont {Allen},\ and\ \citenamefont {Kim}}]{kim}%
  \BibitemOpen
  \bibfield  {author} {\bibinfo {author} {\bibfnamefont {Y.~K.}\ \bibnamefont
  {Kim}}, \bibinfo {author} {\bibfnamefont {O.}~\bibnamefont {Krupin}},
  \bibinfo {author} {\bibfnamefont {J.~D.}\ \bibnamefont {Denlinger}}, \bibinfo
  {author} {\bibfnamefont {A.}~\bibnamefont {Bostwick}}, \bibinfo {author}
  {\bibfnamefont {E.}~\bibnamefont {Rotenberg}}, \bibinfo {author}
  {\bibfnamefont {Q.}~\bibnamefont {Zhao}}, \bibinfo {author} {\bibfnamefont
  {J.~F.}\ \bibnamefont {Mitchell}}, \bibinfo {author} {\bibfnamefont {J.~W.}\
  \bibnamefont {Allen}}, \ and\ \bibinfo {author} {\bibfnamefont {B.~J.}\
  \bibnamefont {Kim}},\ }\href {\doibase 10.1126/science.1251151} {\bibfield
  {journal} {\bibinfo  {journal} {Science}\ } (\bibinfo {year} {2014}),\
  10.1126/science.1251151}\BibitemShut {NoStop}%
\bibitem [{\citenamefont {Iwaya}\ \emph {et~al.}(2007)\citenamefont {Iwaya},
  \citenamefont {Satow}, \citenamefont {Hanaguri}, \citenamefont {Shannon},
  \citenamefont {Yoshida}, \citenamefont {Ikeda}, \citenamefont {He},
  \citenamefont {Kaneko}, \citenamefont {Tokura}, \citenamefont {Yamada} \emph
  {et~al.}}]{iwaya2007local}%
  \BibitemOpen
  \bibfield  {author} {\bibinfo {author} {\bibfnamefont {K.}~\bibnamefont
  {Iwaya}}, \bibinfo {author} {\bibfnamefont {S.}~\bibnamefont {Satow}},
  \bibinfo {author} {\bibfnamefont {T.}~\bibnamefont {Hanaguri}}, \bibinfo
  {author} {\bibfnamefont {N.}~\bibnamefont {Shannon}}, \bibinfo {author}
  {\bibfnamefont {Y.}~\bibnamefont {Yoshida}}, \bibinfo {author} {\bibfnamefont
  {S.}~\bibnamefont {Ikeda}}, \bibinfo {author} {\bibfnamefont
  {J.}~\bibnamefont {He}}, \bibinfo {author} {\bibfnamefont {Y.}~\bibnamefont
  {Kaneko}}, \bibinfo {author} {\bibfnamefont {Y.}~\bibnamefont {Tokura}},
  \bibinfo {author} {\bibfnamefont {T.}~\bibnamefont {Yamada}},  \emph
  {et~al.},\ }\href@noop {} {\bibfield  {journal} {\bibinfo  {journal}
  {Physical review letters}\ }\textbf {\bibinfo {volume} {99}},\ \bibinfo
  {pages} {057208} (\bibinfo {year} {2007})}\BibitemShut {NoStop}%
\bibitem [{\citenamefont {Shimojima}\ \emph {et~al.}(2014)\citenamefont
  {Shimojima}, \citenamefont {Sonobe}, \citenamefont {Malaeb}, \citenamefont
  {Shinada}, \citenamefont {Chainani}, \citenamefont {Shin}, \citenamefont
  {Yoshida}, \citenamefont {Ideta}, \citenamefont {Fujimori}, \citenamefont
  {Kumigashira}, \citenamefont {Ono}, \citenamefont {Nakashima}, \citenamefont
  {Anzai}, \citenamefont {Arita}, \citenamefont {Ino}, \citenamefont
  {Namatame}, \citenamefont {Taniguchi}, \citenamefont {Nakajima},
  \citenamefont {Uchida}, \citenamefont {Tomioka}, \citenamefont {Ito},
  \citenamefont {Kihou}, \citenamefont {Lee}, \citenamefont {Iyo},
  \citenamefont {Eisaki}, \citenamefont {Ohgushi}, \citenamefont {Kasahara},
  \citenamefont {Terashima}, \citenamefont {Ikeda}, \citenamefont {Shibauchi},
  \citenamefont {Matsuda},\ and\ \citenamefont {Ishizaka}}]{ironpnictides_1}%
  \BibitemOpen
  \bibfield  {author} {\bibinfo {author} {\bibfnamefont {T.}~\bibnamefont
  {Shimojima}}, \bibinfo {author} {\bibfnamefont {T.}~\bibnamefont {Sonobe}},
  \bibinfo {author} {\bibfnamefont {W.}~\bibnamefont {Malaeb}}, \bibinfo
  {author} {\bibfnamefont {K.}~\bibnamefont {Shinada}}, \bibinfo {author}
  {\bibfnamefont {A.}~\bibnamefont {Chainani}}, \bibinfo {author}
  {\bibfnamefont {S.}~\bibnamefont {Shin}}, \bibinfo {author} {\bibfnamefont
  {T.}~\bibnamefont {Yoshida}}, \bibinfo {author} {\bibfnamefont
  {S.}~\bibnamefont {Ideta}}, \bibinfo {author} {\bibfnamefont
  {A.}~\bibnamefont {Fujimori}}, \bibinfo {author} {\bibfnamefont
  {H.}~\bibnamefont {Kumigashira}}, \bibinfo {author} {\bibfnamefont
  {K.}~\bibnamefont {Ono}}, \bibinfo {author} {\bibfnamefont {Y.}~\bibnamefont
  {Nakashima}}, \bibinfo {author} {\bibfnamefont {H.}~\bibnamefont {Anzai}},
  \bibinfo {author} {\bibfnamefont {M.}~\bibnamefont {Arita}}, \bibinfo
  {author} {\bibfnamefont {A.}~\bibnamefont {Ino}}, \bibinfo {author}
  {\bibfnamefont {H.}~\bibnamefont {Namatame}}, \bibinfo {author}
  {\bibfnamefont {M.}~\bibnamefont {Taniguchi}}, \bibinfo {author}
  {\bibfnamefont {M.}~\bibnamefont {Nakajima}}, \bibinfo {author}
  {\bibfnamefont {S.}~\bibnamefont {Uchida}}, \bibinfo {author} {\bibfnamefont
  {Y.}~\bibnamefont {Tomioka}}, \bibinfo {author} {\bibfnamefont
  {T.}~\bibnamefont {Ito}}, \bibinfo {author} {\bibfnamefont {K.}~\bibnamefont
  {Kihou}}, \bibinfo {author} {\bibfnamefont {C.~H.}\ \bibnamefont {Lee}},
  \bibinfo {author} {\bibfnamefont {A.}~\bibnamefont {Iyo}}, \bibinfo {author}
  {\bibfnamefont {H.}~\bibnamefont {Eisaki}}, \bibinfo {author} {\bibfnamefont
  {K.}~\bibnamefont {Ohgushi}}, \bibinfo {author} {\bibfnamefont
  {S.}~\bibnamefont {Kasahara}}, \bibinfo {author} {\bibfnamefont
  {T.}~\bibnamefont {Terashima}}, \bibinfo {author} {\bibfnamefont
  {H.}~\bibnamefont {Ikeda}}, \bibinfo {author} {\bibfnamefont
  {T.}~\bibnamefont {Shibauchi}}, \bibinfo {author} {\bibfnamefont
  {Y.}~\bibnamefont {Matsuda}}, \ and\ \bibinfo {author} {\bibfnamefont
  {K.}~\bibnamefont {Ishizaka}},\ }\href {\doibase 10.1103/PhysRevB.89.045101}
  {\bibfield  {journal} {\bibinfo  {journal} {Phys. Rev. B}\ }\textbf {\bibinfo
  {volume} {89}},\ \bibinfo {pages} {045101} (\bibinfo {year}
  {2014})}\BibitemShut {NoStop}%
\bibitem [{\citenamefont {Sato}\ \emph {et~al.}(2008)\citenamefont {Sato},
  \citenamefont {Souma}, \citenamefont {Nakayama}, \citenamefont {Terashima},
  \citenamefont {Sugawara}, \citenamefont {Takahashi}, \citenamefont
  {Kamihara}, \citenamefont {Hirano},\ and\ \citenamefont
  {Hosono}}]{ironbase_1}%
  \BibitemOpen
  \bibfield  {author} {\bibinfo {author} {\bibfnamefont {T.}~\bibnamefont
  {Sato}}, \bibinfo {author} {\bibfnamefont {S.}~\bibnamefont {Souma}},
  \bibinfo {author} {\bibfnamefont {K.}~\bibnamefont {Nakayama}}, \bibinfo
  {author} {\bibfnamefont {K.}~\bibnamefont {Terashima}}, \bibinfo {author}
  {\bibfnamefont {K.}~\bibnamefont {Sugawara}}, \bibinfo {author}
  {\bibfnamefont {T.}~\bibnamefont {Takahashi}}, \bibinfo {author}
  {\bibfnamefont {Y.}~\bibnamefont {Kamihara}}, \bibinfo {author}
  {\bibfnamefont {M.}~\bibnamefont {Hirano}}, \ and\ \bibinfo {author}
  {\bibfnamefont {H.}~\bibnamefont {Hosono}},\ }\href {\doibase
  10.1143/JPSJ.77.063708} {\bibfield  {journal} {\bibinfo  {journal} {Journal
  of the Physical Society of Japan}\ }\textbf {\bibinfo {volume} {77}},\
  \bibinfo {pages} {063708} (\bibinfo {year} {2008})}\BibitemShut {NoStop}%
\bibitem [{\citenamefont {Powell}\ and\ \citenamefont
  {McKenzie}(2011)}]{powell2011quantum}%
  \BibitemOpen
  \bibfield  {author} {\bibinfo {author} {\bibfnamefont {B.}~\bibnamefont
  {Powell}}\ and\ \bibinfo {author} {\bibfnamefont {R.~H.}\ \bibnamefont
  {McKenzie}},\ }\href@noop {} {\bibfield  {journal} {\bibinfo  {journal}
  {Reports on Progress in Physics}\ }\textbf {\bibinfo {volume} {74}},\
  \bibinfo {pages} {056501} (\bibinfo {year} {2011})}\BibitemShut {NoStop}%
\bibitem [{\citenamefont {Chen}\ \emph {et~al.}(2017)\citenamefont {Chen},
  \citenamefont {Pai}, \citenamefont {Chan}, \citenamefont {Takayama},
  \citenamefont {Xu}, \citenamefont {Karn}, \citenamefont {Hasegawa},
  \citenamefont {Chou}, \citenamefont {Mo}, \citenamefont {Fedorov} \emph
  {et~al.}}]{chen2017emergence}%
  \BibitemOpen
  \bibfield  {author} {\bibinfo {author} {\bibfnamefont {P.}~\bibnamefont
  {Chen}}, \bibinfo {author} {\bibfnamefont {W.~W.}\ \bibnamefont {Pai}},
  \bibinfo {author} {\bibfnamefont {Y.-H.}\ \bibnamefont {Chan}}, \bibinfo
  {author} {\bibfnamefont {A.}~\bibnamefont {Takayama}}, \bibinfo {author}
  {\bibfnamefont {C.-Z.}\ \bibnamefont {Xu}}, \bibinfo {author} {\bibfnamefont
  {A.}~\bibnamefont {Karn}}, \bibinfo {author} {\bibfnamefont {S.}~\bibnamefont
  {Hasegawa}}, \bibinfo {author} {\bibfnamefont {M.-Y.}\ \bibnamefont {Chou}},
  \bibinfo {author} {\bibfnamefont {S.-K.}\ \bibnamefont {Mo}}, \bibinfo
  {author} {\bibfnamefont {A.-V.}\ \bibnamefont {Fedorov}},  \emph {et~al.},\
  }\href@noop {} {\bibfield  {journal} {\bibinfo  {journal} {Nature
  communications}\ }\textbf {\bibinfo {volume} {8}},\ \bibinfo {pages} {1}
  (\bibinfo {year} {2017})}\BibitemShut {NoStop}%
\bibitem [{\citenamefont {Borisenko}\ \emph {et~al.}(2008)\citenamefont
  {Borisenko}, \citenamefont {Kordyuk}, \citenamefont {Yaresko}, \citenamefont
  {Zabolotnyy}, \citenamefont {Inosov}, \citenamefont {Schuster}, \citenamefont
  {B{\"u}chner}, \citenamefont {Weber}, \citenamefont {Follath}, \citenamefont
  {Patthey} \emph {et~al.}}]{borisenko2008pseudogap}%
  \BibitemOpen
  \bibfield  {author} {\bibinfo {author} {\bibfnamefont {S.}~\bibnamefont
  {Borisenko}}, \bibinfo {author} {\bibfnamefont {A.}~\bibnamefont {Kordyuk}},
  \bibinfo {author} {\bibfnamefont {A.}~\bibnamefont {Yaresko}}, \bibinfo
  {author} {\bibfnamefont {V.}~\bibnamefont {Zabolotnyy}}, \bibinfo {author}
  {\bibfnamefont {D.}~\bibnamefont {Inosov}}, \bibinfo {author} {\bibfnamefont
  {R.}~\bibnamefont {Schuster}}, \bibinfo {author} {\bibfnamefont
  {B.}~\bibnamefont {B{\"u}chner}}, \bibinfo {author} {\bibfnamefont
  {R.}~\bibnamefont {Weber}}, \bibinfo {author} {\bibfnamefont
  {R.}~\bibnamefont {Follath}}, \bibinfo {author} {\bibfnamefont
  {L.}~\bibnamefont {Patthey}},  \emph {et~al.},\ }\href@noop {} {\bibfield
  {journal} {\bibinfo  {journal} {Physical review letters}\ }\textbf {\bibinfo
  {volume} {100}},\ \bibinfo {pages} {196402} (\bibinfo {year}
  {2008})}\BibitemShut {NoStop}%
\bibitem [{\citenamefont {Varma}\ \emph {et~al.}(1989)\citenamefont {Varma},
  \citenamefont {Littlewood}, \citenamefont {Schmitt-Rink}, \citenamefont
  {Abrahams},\ and\ \citenamefont {Ruckenstein}}]{varma1989phenomenology}%
  \BibitemOpen
  \bibfield  {author} {\bibinfo {author} {\bibfnamefont {C.}~\bibnamefont
  {Varma}}, \bibinfo {author} {\bibfnamefont {P.~B.}\ \bibnamefont
  {Littlewood}}, \bibinfo {author} {\bibfnamefont {S.}~\bibnamefont
  {Schmitt-Rink}}, \bibinfo {author} {\bibfnamefont {E.}~\bibnamefont
  {Abrahams}}, \ and\ \bibinfo {author} {\bibfnamefont {A.}~\bibnamefont
  {Ruckenstein}},\ }\href@noop {} {\bibfield  {journal} {\bibinfo  {journal}
  {Physical Review Letters}\ }\textbf {\bibinfo {volume} {63}},\ \bibinfo
  {pages} {1996} (\bibinfo {year} {1989})}\BibitemShut {NoStop}%
\bibitem [{\citenamefont {Cox}\ and\ \citenamefont
  {Jarrell}(1996)}]{cox1996two}%
  \BibitemOpen
  \bibfield  {author} {\bibinfo {author} {\bibfnamefont {D.}~\bibnamefont
  {Cox}}\ and\ \bibinfo {author} {\bibfnamefont {M.}~\bibnamefont {Jarrell}},\
  }\href@noop {} {\bibfield  {journal} {\bibinfo  {journal} {Journal of
  Physics: Condensed Matter}\ }\textbf {\bibinfo {volume} {8}},\ \bibinfo
  {pages} {9825} (\bibinfo {year} {1996})}\BibitemShut {NoStop}%
\bibitem [{\citenamefont {Lee}\ and\ \citenamefont
  {Phillips}(2012)}]{lee2012non}%
  \BibitemOpen
  \bibfield  {author} {\bibinfo {author} {\bibfnamefont {W.-C.}\ \bibnamefont
  {Lee}}\ and\ \bibinfo {author} {\bibfnamefont {P.~W.}\ \bibnamefont
  {Phillips}},\ }\href@noop {} {\bibfield  {journal} {\bibinfo  {journal}
  {Physical Review B}\ }\textbf {\bibinfo {volume} {86}},\ \bibinfo {pages}
  {245113} (\bibinfo {year} {2012})}\BibitemShut {NoStop}%
\bibitem [{\citenamefont {Iqbal}\ \emph {et~al.}(2012)\citenamefont {Iqbal},
  \citenamefont {Liu},\ and\ \citenamefont {Mezei}}]{iqbal2012lectures}%
  \BibitemOpen
  \bibfield  {author} {\bibinfo {author} {\bibfnamefont {N.}~\bibnamefont
  {Iqbal}}, \bibinfo {author} {\bibfnamefont {H.}~\bibnamefont {Liu}}, \ and\
  \bibinfo {author} {\bibfnamefont {M.}~\bibnamefont {Mezei}},\ }in\ \href@noop
  {} {\emph {\bibinfo {booktitle} {String Theory And Its Applications: TASI
  2010 From meV to the Planck Scale}}}\ (\bibinfo  {publisher} {World
  Scientific},\ \bibinfo {year} {2012})\ pp.\ \bibinfo {pages}
  {707--815}\BibitemShut {NoStop}%
\bibitem [{\citenamefont {Millis}(1993)}]{millis1993effect}%
  \BibitemOpen
  \bibfield  {author} {\bibinfo {author} {\bibfnamefont {A.}~\bibnamefont
  {Millis}},\ }\href@noop {} {\bibfield  {journal} {\bibinfo  {journal}
  {Physical Review B}\ }\textbf {\bibinfo {volume} {48}},\ \bibinfo {pages}
  {7183} (\bibinfo {year} {1993})}\BibitemShut {NoStop}%
\bibitem [{\citenamefont {Castellani}\ \emph {et~al.}(1996)\citenamefont
  {Castellani}, \citenamefont {Di~Castro},\ and\ \citenamefont
  {Grilli}}]{castellani1996non}%
  \BibitemOpen
  \bibfield  {author} {\bibinfo {author} {\bibfnamefont {C.}~\bibnamefont
  {Castellani}}, \bibinfo {author} {\bibfnamefont {C.}~\bibnamefont
  {Di~Castro}}, \ and\ \bibinfo {author} {\bibfnamefont {M.}~\bibnamefont
  {Grilli}},\ }\href@noop {} {\bibfield  {journal} {\bibinfo  {journal}
  {Zeitschrift f{\"u}r Physik B Condensed Matter}\ }\textbf {\bibinfo {volume}
  {103}},\ \bibinfo {pages} {137} (\bibinfo {year} {1996})}\BibitemShut
  {NoStop}%
\bibitem [{\citenamefont {Abanov}\ and\ \citenamefont
  {Chubukov}(2000)}]{abanov2000spin}%
  \BibitemOpen
  \bibfield  {author} {\bibinfo {author} {\bibfnamefont {A.}~\bibnamefont
  {Abanov}}\ and\ \bibinfo {author} {\bibfnamefont {A.~V.}\ \bibnamefont
  {Chubukov}},\ }\href@noop {} {\bibfield  {journal} {\bibinfo  {journal}
  {Physical review letters}\ }\textbf {\bibinfo {volume} {84}},\ \bibinfo
  {pages} {5608} (\bibinfo {year} {2000})}\BibitemShut {NoStop}%
\bibitem [{\citenamefont {Oganesyan}\ \emph {et~al.}(2001)\citenamefont
  {Oganesyan}, \citenamefont {Kivelson},\ and\ \citenamefont
  {Fradkin}}]{oganesyan2001quantum}%
  \BibitemOpen
  \bibfield  {author} {\bibinfo {author} {\bibfnamefont {V.}~\bibnamefont
  {Oganesyan}}, \bibinfo {author} {\bibfnamefont {S.~A.}\ \bibnamefont
  {Kivelson}}, \ and\ \bibinfo {author} {\bibfnamefont {E.}~\bibnamefont
  {Fradkin}},\ }\href@noop {} {\bibfield  {journal} {\bibinfo  {journal}
  {Physical Review B}\ }\textbf {\bibinfo {volume} {64}},\ \bibinfo {pages}
  {195109} (\bibinfo {year} {2001})}\BibitemShut {NoStop}%
\bibitem [{\citenamefont {L{\"o}hneysen}\ \emph {et~al.}(2007)\citenamefont
  {L{\"o}hneysen}, \citenamefont {Rosch}, \citenamefont {Vojta},\ and\
  \citenamefont {W{\"o}lfle}}]{lohneysen2007fermi}%
  \BibitemOpen
  \bibfield  {author} {\bibinfo {author} {\bibfnamefont {H.~v.}\ \bibnamefont
  {L{\"o}hneysen}}, \bibinfo {author} {\bibfnamefont {A.}~\bibnamefont
  {Rosch}}, \bibinfo {author} {\bibfnamefont {M.}~\bibnamefont {Vojta}}, \ and\
  \bibinfo {author} {\bibfnamefont {P.}~\bibnamefont {W{\"o}lfle}},\
  }\href@noop {} {\bibfield  {journal} {\bibinfo  {journal} {Reviews of Modern
  Physics}\ }\textbf {\bibinfo {volume} {79}},\ \bibinfo {pages} {1015}
  (\bibinfo {year} {2007})}\BibitemShut {NoStop}%
\bibitem [{\citenamefont {Metlitski}\ and\ \citenamefont
  {Sachdev}(2010)}]{metlitski2010quantum}%
  \BibitemOpen
  \bibfield  {author} {\bibinfo {author} {\bibfnamefont {M.~A.}\ \bibnamefont
  {Metlitski}}\ and\ \bibinfo {author} {\bibfnamefont {S.}~\bibnamefont
  {Sachdev}},\ }\href@noop {} {\bibfield  {journal} {\bibinfo  {journal}
  {Physical Review B}\ }\textbf {\bibinfo {volume} {82}},\ \bibinfo {pages}
  {075128} (\bibinfo {year} {2010})}\BibitemShut {NoStop}%
\bibitem [{\citenamefont {Abrahams}\ and\ \citenamefont
  {W{\"o}lfle}(2012)}]{abrahams2012critical}%
  \BibitemOpen
  \bibfield  {author} {\bibinfo {author} {\bibfnamefont {E.}~\bibnamefont
  {Abrahams}}\ and\ \bibinfo {author} {\bibfnamefont {P.}~\bibnamefont
  {W{\"o}lfle}},\ }\href@noop {} {\bibfield  {journal} {\bibinfo  {journal}
  {Proceedings of the National Academy of Sciences}\ }\textbf {\bibinfo
  {volume} {109}},\ \bibinfo {pages} {3238} (\bibinfo {year}
  {2012})}\BibitemShut {NoStop}%
\bibitem [{\citenamefont {Isobe}\ \emph {et~al.}(2016)\citenamefont {Isobe},
  \citenamefont {Yang}, \citenamefont {Chubukov}, \citenamefont {Schmalian},\
  and\ \citenamefont {Nagaosa}}]{isobe2016emergent}%
  \BibitemOpen
  \bibfield  {author} {\bibinfo {author} {\bibfnamefont {H.}~\bibnamefont
  {Isobe}}, \bibinfo {author} {\bibfnamefont {B.-J.}\ \bibnamefont {Yang}},
  \bibinfo {author} {\bibfnamefont {A.}~\bibnamefont {Chubukov}}, \bibinfo
  {author} {\bibfnamefont {J.}~\bibnamefont {Schmalian}}, \ and\ \bibinfo
  {author} {\bibfnamefont {N.}~\bibnamefont {Nagaosa}},\ }\href@noop {}
  {\bibfield  {journal} {\bibinfo  {journal} {Physical review letters}\
  }\textbf {\bibinfo {volume} {116}},\ \bibinfo {pages} {076803} (\bibinfo
  {year} {2016})}\BibitemShut {NoStop}%
\bibitem [{\citenamefont {Moriya}(2012)}]{moriya2012spin}%
  \BibitemOpen
  \bibfield  {author} {\bibinfo {author} {\bibfnamefont {T.}~\bibnamefont
  {Moriya}},\ }\href@noop {} {\emph {\bibinfo {title} {Spin fluctuations in
  itinerant electron magnetism}}},\ Vol.~\bibinfo {volume} {56}\ (\bibinfo
  {publisher} {Springer Science \& Business Media},\ \bibinfo {year}
  {2012})\BibitemShut {NoStop}%
\bibitem [{\citenamefont {Liu}\ \emph {et~al.}(2018)\citenamefont {Liu},
  \citenamefont {Xu}, \citenamefont {Qi}, \citenamefont {Sun},\ and\
  \citenamefont {Meng}}]{liu2018itinerant}%
  \BibitemOpen
  \bibfield  {author} {\bibinfo {author} {\bibfnamefont {Z.~H.}\ \bibnamefont
  {Liu}}, \bibinfo {author} {\bibfnamefont {X.~Y.}\ \bibnamefont {Xu}},
  \bibinfo {author} {\bibfnamefont {Y.}~\bibnamefont {Qi}}, \bibinfo {author}
  {\bibfnamefont {K.}~\bibnamefont {Sun}}, \ and\ \bibinfo {author}
  {\bibfnamefont {Z.~Y.}\ \bibnamefont {Meng}},\ }\href@noop {} {\bibfield
  {journal} {\bibinfo  {journal} {Physical Review B}\ }\textbf {\bibinfo
  {volume} {98}},\ \bibinfo {pages} {045116} (\bibinfo {year}
  {2018})}\BibitemShut {NoStop}%
\bibitem [{\citenamefont {Lee}(2018)}]{lee2018recent}%
  \BibitemOpen
  \bibfield  {author} {\bibinfo {author} {\bibfnamefont {S.-S.}\ \bibnamefont
  {Lee}},\ }\href@noop {} {\bibfield  {journal} {\bibinfo  {journal} {Annual
  Review of Condensed Matter Physics}\ }\textbf {\bibinfo {volume} {9}},\
  \bibinfo {pages} {227} (\bibinfo {year} {2018})}\BibitemShut {NoStop}%
\bibitem [{\citenamefont {Loktev}\ \emph {et~al.}(2000)\citenamefont {Loktev},
  \citenamefont {Sharapov}, \citenamefont {Quick},\ and\ \citenamefont
  {Sharapov}}]{loktev2000phase}%
  \BibitemOpen
  \bibfield  {author} {\bibinfo {author} {\bibfnamefont {V.~M.}\ \bibnamefont
  {Loktev}}, \bibinfo {author} {\bibfnamefont {S.~G.}\ \bibnamefont
  {Sharapov}}, \bibinfo {author} {\bibfnamefont {R.~M.}\ \bibnamefont {Quick}},
  \ and\ \bibinfo {author} {\bibfnamefont {S.}~\bibnamefont {Sharapov}},\
  }\href@noop {} {\bibfield  {journal} {\bibinfo  {journal} {Low Temperature
  Physics}\ }\textbf {\bibinfo {volume} {26}},\ \bibinfo {pages} {414}
  (\bibinfo {year} {2000})}\BibitemShut {NoStop}%
\bibitem [{\citenamefont {Loktev}\ \emph {et~al.}(2001)\citenamefont {Loktev},
  \citenamefont {Quick},\ and\ \citenamefont {Sharapov}}]{loktev2001phase}%
  \BibitemOpen
  \bibfield  {author} {\bibinfo {author} {\bibfnamefont {V.~M.}\ \bibnamefont
  {Loktev}}, \bibinfo {author} {\bibfnamefont {R.~M.}\ \bibnamefont {Quick}}, \
  and\ \bibinfo {author} {\bibfnamefont {S.~G.}\ \bibnamefont {Sharapov}},\
  }\href@noop {} {\bibfield  {journal} {\bibinfo  {journal} {Physics Reports}\
  }\textbf {\bibinfo {volume} {349}},\ \bibinfo {pages} {1} (\bibinfo {year}
  {2001})}\BibitemShut {NoStop}%
\bibitem [{\citenamefont {Schmalian}\ \emph {et~al.}(1999)\citenamefont
  {Schmalian}, \citenamefont {Pines},\ and\ \citenamefont
  {Stojkovi{\'c}}}]{schmalian1999microscopic}%
  \BibitemOpen
  \bibfield  {author} {\bibinfo {author} {\bibfnamefont {J.}~\bibnamefont
  {Schmalian}}, \bibinfo {author} {\bibfnamefont {D.}~\bibnamefont {Pines}}, \
  and\ \bibinfo {author} {\bibfnamefont {B.}~\bibnamefont {Stojkovi{\'c}}},\
  }\href@noop {} {\bibfield  {journal} {\bibinfo  {journal} {Physical Review
  B}\ }\textbf {\bibinfo {volume} {60}},\ \bibinfo {pages} {667} (\bibinfo
  {year} {1999})}\BibitemShut {NoStop}%
\bibitem [{\citenamefont {Kuchinskii}\ and\ \citenamefont
  {Sadovskii}(1999)}]{sdw}%
  \BibitemOpen
  \bibfield  {author} {\bibinfo {author} {\bibfnamefont {E.}~\bibnamefont
  {Kuchinskii}}\ and\ \bibinfo {author} {\bibfnamefont {M.}~\bibnamefont
  {Sadovskii}},\ }\href@noop {} {\bibfield  {journal} {\bibinfo  {journal}
  {Journal of Experimental and Theoretical Physics}\ }\textbf {\bibinfo
  {volume} {88}},\ \bibinfo {pages} {968} (\bibinfo {year} {1999})}\BibitemShut
  {NoStop}%
\bibitem [{\citenamefont {Sedrakyan}\ and\ \citenamefont
  {Chubukov}(2010)}]{sedrakyan2010pseudogap}%
  \BibitemOpen
  \bibfield  {author} {\bibinfo {author} {\bibfnamefont {T.~A.}\ \bibnamefont
  {Sedrakyan}}\ and\ \bibinfo {author} {\bibfnamefont {A.~V.}\ \bibnamefont
  {Chubukov}},\ }\href@noop {} {\bibfield  {journal} {\bibinfo  {journal}
  {Physical Review B}\ }\textbf {\bibinfo {volume} {81}},\ \bibinfo {pages}
  {174536} (\bibinfo {year} {2010})}\BibitemShut {NoStop}%
\bibitem [{\citenamefont {Lee}\ \emph {et~al.}(1973)\citenamefont {Lee},
  \citenamefont {Rice},\ and\ \citenamefont {Anderson}}]{PhysRevLett.31.462}%
  \BibitemOpen
  \bibfield  {author} {\bibinfo {author} {\bibfnamefont {P.~A.}\ \bibnamefont
  {Lee}}, \bibinfo {author} {\bibfnamefont {T.~M.}\ \bibnamefont {Rice}}, \
  and\ \bibinfo {author} {\bibfnamefont {P.~W.}\ \bibnamefont {Anderson}},\
  }\href {\doibase 10.1103/PhysRevLett.31.462} {\bibfield  {journal} {\bibinfo
  {journal} {Phys. Rev. Lett.}\ }\textbf {\bibinfo {volume} {31}},\ \bibinfo
  {pages} {462} (\bibinfo {year} {1973})}\BibitemShut {NoStop}%
\bibitem [{\citenamefont {Kuchinskii}\ \emph {et~al.}(2012)\citenamefont
  {Kuchinskii}, \citenamefont {Nekrasov},\ and\ \citenamefont
  {Sadovskii}}]{kuchinskii2012electronic}%
  \BibitemOpen
  \bibfield  {author} {\bibinfo {author} {\bibfnamefont {E.}~\bibnamefont
  {Kuchinskii}}, \bibinfo {author} {\bibfnamefont {I.}~\bibnamefont
  {Nekrasov}}, \ and\ \bibinfo {author} {\bibfnamefont {M.}~\bibnamefont
  {Sadovskii}},\ }\href@noop {} {\bibfield  {journal} {\bibinfo  {journal}
  {Journal of Experimental and Theoretical Physics}\ }\textbf {\bibinfo
  {volume} {114}},\ \bibinfo {pages} {671} (\bibinfo {year}
  {2012})}\BibitemShut {NoStop}%
\bibitem [{\citenamefont {Yamase}\ and\ \citenamefont
  {Metzner}(2012)}]{PhysRevLett.108.186405}%
  \BibitemOpen
  \bibfield  {author} {\bibinfo {author} {\bibfnamefont {H.}~\bibnamefont
  {Yamase}}\ and\ \bibinfo {author} {\bibfnamefont {W.}~\bibnamefont
  {Metzner}},\ }\href {\doibase 10.1103/PhysRevLett.108.186405} {\bibfield
  {journal} {\bibinfo  {journal} {Phys. Rev. Lett.}\ }\textbf {\bibinfo
  {volume} {108}},\ \bibinfo {pages} {186405} (\bibinfo {year}
  {2012})}\BibitemShut {NoStop}%
\bibitem [{\citenamefont {Bejas}\ \emph {et~al.}(2011)\citenamefont {Bejas},
  \citenamefont {Buzon}, \citenamefont {Greco},\ and\ \citenamefont
  {Foussats}}]{PhysRevB.83.014514}%
  \BibitemOpen
  \bibfield  {author} {\bibinfo {author} {\bibfnamefont {M.}~\bibnamefont
  {Bejas}}, \bibinfo {author} {\bibfnamefont {G.}~\bibnamefont {Buzon}},
  \bibinfo {author} {\bibfnamefont {A.}~\bibnamefont {Greco}}, \ and\ \bibinfo
  {author} {\bibfnamefont {A.}~\bibnamefont {Foussats}},\ }\href {\doibase
  10.1103/PhysRevB.83.014514} {\bibfield  {journal} {\bibinfo  {journal} {Phys.
  Rev. B}\ }\textbf {\bibinfo {volume} {83}},\ \bibinfo {pages} {014514}
  (\bibinfo {year} {2011})}\BibitemShut {NoStop}%
\bibitem [{\citenamefont {Kordyuk}(2015)}]{kordyuk2015pseudogap}%
  \BibitemOpen
  \bibfield  {author} {\bibinfo {author} {\bibfnamefont {A.}~\bibnamefont
  {Kordyuk}},\ }\href@noop {} {\bibfield  {journal} {\bibinfo  {journal} {Low
  Temperature Physics}\ }\textbf {\bibinfo {volume} {41}},\ \bibinfo {pages}
  {319} (\bibinfo {year} {2015})}\BibitemShut {NoStop}%
\bibitem [{\citenamefont {Kuleeva}\ \emph {et~al.}(2014)\citenamefont
  {Kuleeva}, \citenamefont {Kuchinskii},\ and\ \citenamefont
  {Sadovskii}}]{kuleeva2014normal}%
  \BibitemOpen
  \bibfield  {author} {\bibinfo {author} {\bibfnamefont {N.}~\bibnamefont
  {Kuleeva}}, \bibinfo {author} {\bibfnamefont {E.}~\bibnamefont {Kuchinskii}},
  \ and\ \bibinfo {author} {\bibfnamefont {M.}~\bibnamefont {Sadovskii}},\
  }\href@noop {} {\bibfield  {journal} {\bibinfo  {journal} {Journal of
  Experimental and Theoretical Physics}\ }\textbf {\bibinfo {volume} {119}},\
  \bibinfo {pages} {264} (\bibinfo {year} {2014})}\BibitemShut {NoStop}%
\bibitem [{\citenamefont {KOHN}\ and\ \citenamefont
  {SHERRINGTON}(1970)}]{KohnBosons}%
  \BibitemOpen
  \bibfield  {author} {\bibinfo {author} {\bibfnamefont {W.}~\bibnamefont
  {KOHN}}\ and\ \bibinfo {author} {\bibfnamefont {D.}~\bibnamefont
  {SHERRINGTON}},\ }\href {\doibase 10.1103/RevModPhys.42.1} {\bibfield
  {journal} {\bibinfo  {journal} {Rev. Mod. Phys.}\ }\textbf {\bibinfo {volume}
  {42}},\ \bibinfo {pages} {1} (\bibinfo {year} {1970})}\BibitemShut {NoStop}%
\bibitem [{\citenamefont {Yildirim}\ \emph {et~al.}(2011)\citenamefont
  {Yildirim}, \citenamefont {Ku} \emph {et~al.}}]{yildirim2011kinetics}%
  \BibitemOpen
  \bibfield  {author} {\bibinfo {author} {\bibfnamefont {Y.}~\bibnamefont
  {Yildirim}}, \bibinfo {author} {\bibfnamefont {W.}~\bibnamefont {Ku}},  \emph
  {et~al.},\ }\href@noop {} {\bibfield  {journal} {\bibinfo  {journal}
  {Physical Review X}\ }\textbf {\bibinfo {volume} {1}},\ \bibinfo {pages}
  {011011} (\bibinfo {year} {2011})}\BibitemShut {NoStop}%
\bibitem [{Note1()}]{Note1}%
  \BibitemOpen
  \bibinfo {note} {Figure \ref {Fig_1}(d) already resembles the experimental
  Fermi arc, although there is some debate~\cite {yang2011reconstructed} on the
  shape at the tip of the arc. In our theory, this inessential detail is
  strongly dependent on the choice of bare fermionic structure, and such
  features could easily be accounted for by using a more 'first-principles'
  system incorporating, for example, strong magnetic correlations of higher
  energy scale.}\BibitemShut {Stop}%
\bibitem [{\citenamefont {Jiang}\ \emph {et~al.}(2019)\citenamefont {Jiang},
  \citenamefont {Zou}, \citenamefont {Ku} \emph {et~al.}}]{jiang2019non}%
  \BibitemOpen
  \bibfield  {author} {\bibinfo {author} {\bibfnamefont {S.}~\bibnamefont
  {Jiang}}, \bibinfo {author} {\bibfnamefont {L.}~\bibnamefont {Zou}}, \bibinfo
  {author} {\bibfnamefont {W.}~\bibnamefont {Ku}},  \emph {et~al.},\
  }\href@noop {} {\bibfield  {journal} {\bibinfo  {journal} {Physical Review
  B}\ }\textbf {\bibinfo {volume} {99}},\ \bibinfo {pages} {104507} (\bibinfo
  {year} {2019})}\BibitemShut {NoStop}%
\bibitem [{\citenamefont {Zhang}\ and\ \citenamefont
  {Chen}(2021)}]{zhang2021infinity}%
  \BibitemOpen
  \bibfield  {author} {\bibinfo {author} {\bibfnamefont {X.-T.}\ \bibnamefont
  {Zhang}}\ and\ \bibinfo {author} {\bibfnamefont {G.}~\bibnamefont {Chen}},\
  }\href@noop {} {\bibfield  {journal} {\bibinfo  {journal} {arXiv preprint
  arXiv:2102.09272}\ } (\bibinfo {year} {2021})}\BibitemShut {NoStop}%
\bibitem [{\citenamefont {Lake}\ \emph {et~al.}(2021)\citenamefont {Lake},
  \citenamefont {Senthil},\ and\ \citenamefont {Vishwanath}}]{lake2021bose}%
  \BibitemOpen
  \bibfield  {author} {\bibinfo {author} {\bibfnamefont {E.}~\bibnamefont
  {Lake}}, \bibinfo {author} {\bibfnamefont {T.}~\bibnamefont {Senthil}}, \
  and\ \bibinfo {author} {\bibfnamefont {A.}~\bibnamefont {Vishwanath}},\
  }\href@noop {} {\bibfield  {journal} {\bibinfo  {journal} {arXiv preprint
  arXiv:2101.02197}\ } (\bibinfo {year} {2021})}\BibitemShut {NoStop}%
\bibitem [{\citenamefont {Leong}\ \emph {et~al.}(2017)\citenamefont {Leong},
  \citenamefont {Setty}, \citenamefont {Limtragool},\ and\ \citenamefont
  {Phillips}}]{leong2017power}%
  \BibitemOpen
  \bibfield  {author} {\bibinfo {author} {\bibfnamefont {Z.}~\bibnamefont
  {Leong}}, \bibinfo {author} {\bibfnamefont {C.}~\bibnamefont {Setty}},
  \bibinfo {author} {\bibfnamefont {K.}~\bibnamefont {Limtragool}}, \ and\
  \bibinfo {author} {\bibfnamefont {P.~W.}\ \bibnamefont {Phillips}},\
  }\href@noop {} {\bibfield  {journal} {\bibinfo  {journal} {Physical Review
  B}\ }\textbf {\bibinfo {volume} {96}},\ \bibinfo {pages} {205101} (\bibinfo
  {year} {2017})}\BibitemShut {NoStop}%
\bibitem [{\citenamefont {Abrahams}\ and\ \citenamefont
  {Varma}(2000)}]{abrahams2000angle}%
  \BibitemOpen
  \bibfield  {author} {\bibinfo {author} {\bibfnamefont {E.}~\bibnamefont
  {Abrahams}}\ and\ \bibinfo {author} {\bibfnamefont {C.}~\bibnamefont
  {Varma}},\ }\href@noop {} {\bibfield  {journal} {\bibinfo  {journal}
  {Proceedings of the National Academy of Sciences}\ }\textbf {\bibinfo
  {volume} {97}},\ \bibinfo {pages} {5714} (\bibinfo {year}
  {2000})}\BibitemShut {NoStop}%
\bibitem [{\citenamefont {Dai}\ \emph {et~al.}(2020)\citenamefont {Dai},
  \citenamefont {Senthil},\ and\ \citenamefont {Lee}}]{dai2020modeling}%
  \BibitemOpen
  \bibfield  {author} {\bibinfo {author} {\bibfnamefont {Z.}~\bibnamefont
  {Dai}}, \bibinfo {author} {\bibfnamefont {T.}~\bibnamefont {Senthil}}, \ and\
  \bibinfo {author} {\bibfnamefont {P.~A.}\ \bibnamefont {Lee}},\ }\href@noop
  {} {\bibfield  {journal} {\bibinfo  {journal} {Physical Review B}\ }\textbf
  {\bibinfo {volume} {101}},\ \bibinfo {pages} {064502} (\bibinfo {year}
  {2020})}\BibitemShut {NoStop}%
\bibitem [{\citenamefont {Yildirim}\ \emph {et~al.}(2015)\citenamefont
  {Yildirim}, \citenamefont {Ku} \emph {et~al.}}]{yildirim2015weak}%
  \BibitemOpen
  \bibfield  {author} {\bibinfo {author} {\bibfnamefont {Y.}~\bibnamefont
  {Yildirim}}, \bibinfo {author} {\bibfnamefont {W.}~\bibnamefont {Ku}},  \emph
  {et~al.},\ }\href@noop {} {\bibfield  {journal} {\bibinfo  {journal}
  {Physical Review B}\ }\textbf {\bibinfo {volume} {92}},\ \bibinfo {pages}
  {180501} (\bibinfo {year} {2015})}\BibitemShut {NoStop}%
\bibitem [{\citenamefont {Lang}\ \emph {et~al.}(2019)\citenamefont {Lang},
  \citenamefont {Yang},\ and\ \citenamefont {Ku}}]{lang2019mottness}%
  \BibitemOpen
  \bibfield  {author} {\bibinfo {author} {\bibfnamefont {Z.-J.}\ \bibnamefont
  {Lang}}, \bibinfo {author} {\bibfnamefont {F.}~\bibnamefont {Yang}}, \ and\
  \bibinfo {author} {\bibfnamefont {W.}~\bibnamefont {Ku}},\ }\href@noop {}
  {\bibfield  {journal} {\bibinfo  {journal} {arXiv preprint arXiv:1902.11206}\
  } (\bibinfo {year} {2019})}\BibitemShut {NoStop}%
\bibitem [{\citenamefont {Zeng}\ \emph {et~al.}(2021)\citenamefont {Zeng},
  \citenamefont {Hegg}, \citenamefont {Zou}, \citenamefont {Jiang},\ and\
  \citenamefont {Ku}}]{zeng2021transport}%
  \BibitemOpen
  \bibfield  {author} {\bibinfo {author} {\bibfnamefont {T.}~\bibnamefont
  {Zeng}}, \bibinfo {author} {\bibfnamefont {A.}~\bibnamefont {Hegg}}, \bibinfo
  {author} {\bibfnamefont {L.}~\bibnamefont {Zou}}, \bibinfo {author}
  {\bibfnamefont {S.}~\bibnamefont {Jiang}}, \ and\ \bibinfo {author}
  {\bibfnamefont {W.}~\bibnamefont {Ku}},\ }\href@noop {} {\bibfield  {journal}
  {\bibinfo  {journal} {arXiv preprint arXiv:2112.05747}\ } (\bibinfo {year}
  {2021})}\BibitemShut {NoStop}%
\bibitem [{\citenamefont {Lang}\ \emph {et~al.}(2022)\citenamefont {Lang},
  \citenamefont {Yang},\ and\ \citenamefont {Ku}}]{lang2022mottness}%
  \BibitemOpen
  \bibfield  {author} {\bibinfo {author} {\bibfnamefont {Z.-J.}\ \bibnamefont
  {Lang}}, \bibinfo {author} {\bibfnamefont {F.}~\bibnamefont {Yang}}, \ and\
  \bibinfo {author} {\bibfnamefont {W.}~\bibnamefont {Ku}},\ }\href@noop {}
  {\bibfield  {journal} {\bibinfo  {journal} {New Journal of Physics}\ }\textbf
  {\bibinfo {volume} {24}},\ \bibinfo {pages} {093026} (\bibinfo {year}
  {2022})}\BibitemShut {NoStop}%
\bibitem [{\citenamefont {Alexandrov}\ and\ \citenamefont
  {Mott}(1994)}]{alexandrov1994bipolarons}%
  \BibitemOpen
  \bibfield  {author} {\bibinfo {author} {\bibfnamefont {A.}~\bibnamefont
  {Alexandrov}}\ and\ \bibinfo {author} {\bibfnamefont {N.}~\bibnamefont
  {Mott}},\ }\href@noop {} {\bibfield  {journal} {\bibinfo  {journal} {Reports
  on Progress in Physics}\ }\textbf {\bibinfo {volume} {57}},\ \bibinfo {pages}
  {1197} (\bibinfo {year} {1994})}\BibitemShut {NoStop}%
\bibitem [{\citenamefont {Zhang}\ \emph {et~al.}(2023)\citenamefont {Zhang},
  \citenamefont {Sous}, \citenamefont {Reichman}, \citenamefont {Berciu},
  \citenamefont {Millis}, \citenamefont {Prokof'ev},\ and\ \citenamefont
  {Svistunov}}]{PhysRevX.13.011010}%
  \BibitemOpen
  \bibfield  {author} {\bibinfo {author} {\bibfnamefont {C.}~\bibnamefont
  {Zhang}}, \bibinfo {author} {\bibfnamefont {J.}~\bibnamefont {Sous}},
  \bibinfo {author} {\bibfnamefont {D.~R.}\ \bibnamefont {Reichman}}, \bibinfo
  {author} {\bibfnamefont {M.}~\bibnamefont {Berciu}}, \bibinfo {author}
  {\bibfnamefont {A.~J.}\ \bibnamefont {Millis}}, \bibinfo {author}
  {\bibfnamefont {N.~V.}\ \bibnamefont {Prokof'ev}}, \ and\ \bibinfo {author}
  {\bibfnamefont {B.~V.}\ \bibnamefont {Svistunov}},\ }\href {\doibase
  10.1103/PhysRevX.13.011010} {\bibfield  {journal} {\bibinfo  {journal} {Phys.
  Rev. X}\ }\textbf {\bibinfo {volume} {13}},\ \bibinfo {pages} {011010}
  (\bibinfo {year} {2023})}\BibitemShut {NoStop}%
\bibitem [{\citenamefont {Lau}\ \emph {et~al.}(2011)\citenamefont {Lau},
  \citenamefont {Berciu},\ and\ \citenamefont
  {Sawatzky}}]{PhysRevLett.106.036401}%
  \BibitemOpen
  \bibfield  {author} {\bibinfo {author} {\bibfnamefont {B.}~\bibnamefont
  {Lau}}, \bibinfo {author} {\bibfnamefont {M.}~\bibnamefont {Berciu}}, \ and\
  \bibinfo {author} {\bibfnamefont {G.~A.}\ \bibnamefont {Sawatzky}},\ }\href
  {\doibase 10.1103/PhysRevLett.106.036401} {\bibfield  {journal} {\bibinfo
  {journal} {Phys. Rev. Lett.}\ }\textbf {\bibinfo {volume} {106}},\ \bibinfo
  {pages} {036401} (\bibinfo {year} {2011})}\BibitemShut {NoStop}%
\bibitem [{\citenamefont {White}\ and\ \citenamefont
  {Scalapino}(1997)}]{PhysRevB.55.6504}%
  \BibitemOpen
  \bibfield  {author} {\bibinfo {author} {\bibfnamefont {S.~R.}\ \bibnamefont
  {White}}\ and\ \bibinfo {author} {\bibfnamefont {D.~J.}\ \bibnamefont
  {Scalapino}},\ }\href {\doibase 10.1103/PhysRevB.55.6504} {\bibfield
  {journal} {\bibinfo  {journal} {Phys. Rev. B}\ }\textbf {\bibinfo {volume}
  {55}},\ \bibinfo {pages} {6504} (\bibinfo {year} {1997})}\BibitemShut
  {NoStop}%
\bibitem [{\citenamefont {Altman}\ and\ \citenamefont
  {Auerbach}(2002)}]{altman2002plaquette}%
  \BibitemOpen
  \bibfield  {author} {\bibinfo {author} {\bibfnamefont {E.}~\bibnamefont
  {Altman}}\ and\ \bibinfo {author} {\bibfnamefont {A.}~\bibnamefont
  {Auerbach}},\ }\href@noop {} {\bibfield  {journal} {\bibinfo  {journal}
  {Physical Review B}\ }\textbf {\bibinfo {volume} {65}},\ \bibinfo {pages}
  {104508} (\bibinfo {year} {2002})}\BibitemShut {NoStop}%
\bibitem [{\citenamefont {Sun}\ \emph {et~al.}(2019)\citenamefont {Sun},
  \citenamefont {Zhu},\ and\ \citenamefont {Weng}}]{PhysRevLett.123.016601}%
  \BibitemOpen
  \bibfield  {author} {\bibinfo {author} {\bibfnamefont {R.-Y.}\ \bibnamefont
  {Sun}}, \bibinfo {author} {\bibfnamefont {Z.}~\bibnamefont {Zhu}}, \ and\
  \bibinfo {author} {\bibfnamefont {Z.-Y.}\ \bibnamefont {Weng}},\ }\href
  {\doibase 10.1103/PhysRevLett.123.016601} {\bibfield  {journal} {\bibinfo
  {journal} {Phys. Rev. Lett.}\ }\textbf {\bibinfo {volume} {123}},\ \bibinfo
  {pages} {016601} (\bibinfo {year} {2019})}\BibitemShut {NoStop}%
\bibitem [{\citenamefont {Zhao}\ \emph {et~al.}(2022)\citenamefont {Zhao},
  \citenamefont {Chen}, \citenamefont {Zhang},\ and\ \citenamefont
  {Weng}}]{zhao2022two}%
  \BibitemOpen
  \bibfield  {author} {\bibinfo {author} {\bibfnamefont {J.-Y.}\ \bibnamefont
  {Zhao}}, \bibinfo {author} {\bibfnamefont {S.~A.}\ \bibnamefont {Chen}},
  \bibinfo {author} {\bibfnamefont {H.-K.}\ \bibnamefont {Zhang}}, \ and\
  \bibinfo {author} {\bibfnamefont {Z.-Y.}\ \bibnamefont {Weng}},\ }\href@noop
  {} {\bibfield  {journal} {\bibinfo  {journal} {Physical Review X}\ }\textbf
  {\bibinfo {volume} {12}},\ \bibinfo {pages} {011062} (\bibinfo {year}
  {2022})}\BibitemShut {NoStop}%
\bibitem [{\citenamefont {Zhang}\ and\ \citenamefont
  {Rice}(1988)}]{zhang1988effective}%
  \BibitemOpen
  \bibfield  {author} {\bibinfo {author} {\bibfnamefont {F.}~\bibnamefont
  {Zhang}}\ and\ \bibinfo {author} {\bibfnamefont {T.}~\bibnamefont {Rice}},\
  }\href@noop {} {\bibfield  {journal} {\bibinfo  {journal} {Physical Review
  B}\ }\textbf {\bibinfo {volume} {37}},\ \bibinfo {pages} {3759} (\bibinfo
  {year} {1988})}\BibitemShut {NoStop}%
\bibitem [{\citenamefont {Yang}\ \emph {et~al.}(2011)\citenamefont {Yang},
  \citenamefont {Rameau}, \citenamefont {Pan}, \citenamefont {Gu},
  \citenamefont {Johnson}, \citenamefont {Claus}, \citenamefont {Hinks},\ and\
  \citenamefont {Kidd}}]{yang2011reconstructed}%
  \BibitemOpen
  \bibfield  {author} {\bibinfo {author} {\bibfnamefont {H.-B.}\ \bibnamefont
  {Yang}}, \bibinfo {author} {\bibfnamefont {J.}~\bibnamefont {Rameau}},
  \bibinfo {author} {\bibfnamefont {Z.-H.}\ \bibnamefont {Pan}}, \bibinfo
  {author} {\bibfnamefont {G.}~\bibnamefont {Gu}}, \bibinfo {author}
  {\bibfnamefont {P.}~\bibnamefont {Johnson}}, \bibinfo {author} {\bibfnamefont
  {H.}~\bibnamefont {Claus}}, \bibinfo {author} {\bibfnamefont
  {D.}~\bibnamefont {Hinks}}, \ and\ \bibinfo {author} {\bibfnamefont
  {T.}~\bibnamefont {Kidd}},\ }\href@noop {} {\bibfield  {journal} {\bibinfo
  {journal} {Physical review letters}\ }\textbf {\bibinfo {volume} {107}},\
  \bibinfo {pages} {047003} (\bibinfo {year} {2011})}\BibitemShut {NoStop}%
\end{thebibliography}%
\clearpage
\end{document}